\documentclass[prb,twocolumn,showpacs]{revtex4-1}
\usepackage[latin1]{inputenc}
\usepackage{graphicx}
\usepackage{amssymb}
\usepackage{amsmath}
\usepackage{xspace}
\usepackage{dcolumn}
\usepackage{bm}
\usepackage{color}
\usepackage{float}
\usepackage[extra]{tipa}

\setcitestyle{square,numbers}

\newfont{\tensy}{cmsy10}
\newcommand{\chem}[1]{{$\fontdimen16\tensy=3.0pt
    \fontdimen17\tensy=3.0pt \mathrm{#1}$}}

\newcommand{\eg}[0]{e.g.\@\xspace}

\newcommand{\UP}[0]{\uparrow}
\newcommand{\DO}[0]{\downarrow}

\newcommand{\oP}{\hat{P}}
\newcommand{\oQ}{\hat{Q}}

\newcommand{\on}{\hat{n}}

\newcommand{\rmd}{\text{d}}

\newcommand{\bit}{\begin{itemize}}
\newcommand{\eit}{\end{itemize}}

\newcommand{\oh}{\mbox{$\frac{1}{2}$}}

\newcommand{\om}[0]{\omega}

\newcommand{\kF}{k_\text{F}}

\newcommand{\nag}{{\phantom{\dag}}}

\newcommand{\psib}{\overline{\psi}}
\newcommand{\phib}{\overline{\phi}}

\newcommand{\Kr}{K_\rho}
\newcommand{\Ks}{K_\sigma}

\newcommand{\las}[0]{\langle}
\newcommand{\ras}[0]{\rangle}

\newcommand{\la}[0]{\left\las}
\newcommand{\ra}[0]{\right\ras}
\newcommand{\ket}[1]{\left|#1\ra}
\newcommand{\bra}[1]{\la#1\right|}

\begin{document}


\title{Excitation spectra and correlation functions of quantum Su-Schrieffer-Heeger models}

\author{Manuel Weber}

\author{Fakher F. Assaad}

\author{Martin Hohenadler}

\affiliation{\mbox{Institut f\"ur Theoretische Physik und Astrophysik,
    Universit\"at W\"urzburg, 97074 W\"urzburg, Germany}}

\begin{abstract}
  We study one-dimensional Su-Schrieffer-Heeger (SSH) models with quantum
  phonons using a continuous-time quantum Monte Carlo method. Within
  statistical errors, we obtain identical results for the SSH model with
  acoustic phonons, and a related model with a coupling to an optical bond
  phonon mode. Based on this agreement, we first study the Peierls
  metal-insulator transition of the spinless SSH model, and relate it 
  to the Kosterlitz-Thouless transition of a spinless Luttinger
  liquid. In the Peierls phase, the spectral functions reveal the
  single-particle and charge gap, and a central peak related to long-range
  order. For the spinful SSH model, which has a dimerized ground state for any
  nonzero coupling, we reveal a symmetry-related degeneracy of spin and
  charge excitations, and the expected spin and charge gaps as well as a
  central peak. Finally, we study the SSH-$UV$ model with  electron-phonon and
  electron-electron interaction. We observe a Mott phase with 
  critical spin and bond correlations at weak electron-phonon coupling, and a
  Peierls phase with gapped spin excitations at strong coupling. We relate
  our findings to the extended Hubbard model, and discuss the physical origin of the
  agreement between optical and acoustic phonons.
\end{abstract}

\date{\today}


\maketitle

\section{Introduction}\label{sec:intro}

In 1979, Su, Schrieffer and Heeger (SSH) introduced a simple microscopic
model of polyacetylene with a modulation of the electronic hopping as a
result of distortions of the carbon bonds \cite{PhysRevLett.42.1698}. It has
since gained a much wider relevance due to the fact that it supports
topological excitations (solitons) with fractional charge
\cite{PhysRevD.13.3398,PhysRevLett.42.1698,RevModPhys.60.781}, and its
relation to one-dimensional topological insulators \cite{RevModPhys.83.1057}
in the BDI class \cite{PhysRevB.78.195125}.  As a model for correlated
electrons, the case of a half-filled band has attracted particular interest,
because it supports a Peierls and (in the presence of Coulomb interaction) a
Mott insulating phase. If the lattice is treated quantum mechanically, the
model constitutes a rich and complex many-body problem.

In the adiabatic limit (or at the mean-field level), a dimerized Peierls
state results at $T=0$ for any nonzero electron-phonon coupling
\cite{PhysRevLett.42.1698,PhysRevB.27.1680}. Although a classical treatment
of the lattice (see, \eg,
Refs.~\cite{PhysRevB.57.11838,PhysRevB.87.245116,deRaedtSSH96,PhysRevB.87.245116})
appears justified for polyacetylene given the large oscillator mass
\cite{PhysRevLett.42.1698}, quantum fluctuations of the lattice cause a
significant reduction of the dimerization
\cite{Su1982497,PhysRevB.25.7789,PhysRevB.27.1680,PhysRevB.33.5141,PhysRevB.54.7965}
and affect the low-lying excitations \cite{PhysRevB.65.075107}.  Quite
surprisingly, numerical \cite{Su1982497,PhysRevB.27.1680,PhysRevB.73.045106}
and analytical
\cite{PhysRevB.27.1680,PhysRevLett.60.2089,Ba.Bo.07,Bakrim2015} results for
the SSH model without electron-electron interaction suggest that lattice
fluctuations do not destroy the Peierls state, in contrast to the spinless
SSH model \cite{PhysRevB.27.1680} or Holstein models
\cite{Hirsch83a,BuMKHa98,JeZhWh99}. Interestingly, the Peierls state does
become unstable with respect to quantum fluctuations if additional Coulomb
repulsion between electrons is taken into account, as appropriate for
polyacetylene \cite{Baeriswylreview92}. In fact, the single-particle gap
found experimentally has been argued to be predominantly a result of
electronic correlations, with only a small component from the dimerization of
the lattice \cite{PhysRevB.57.11838}. A finite phonon frequency may also
crucially affect the balance between interactions
\cite{PhysRevLett.51.296,PhysRevLett.62.1053}.  The SSH models are also
related to spin-Peierls models
\cite{PhysRevB.27.1680,PhysRevB.67.245103}, studied in detail to understand
the physics of \chem{CuGeO_3} \cite{PhysRevLett.70.3651}, see for example
Refs.~\cite{PhysRevLett.81.3956,PhysRevLett.83.195}.  For a more complete
overview of previous work on polyacetylene we refer to
Refs.~\cite{Baeriswylreview92,Barfordbook}.

The momentum dependence of the phonon dispersion and the interaction make
numerical studies of the SSH model more challenging than for models with a
Holstein interaction \cite{Ho59a}. Therefore, several open problems remain,
in particular the calculation of momentum-resolved excitation
spectra. Besides, some of the recent numerical work has focused on a
simplified model with optical phonons and a different interaction term
\cite{PhysRevB.67.245103}. While theory suggests that the low-energy physics
depends solely on the phonon frequency at $q=2\kF$
\cite{PhysRevB.27.1680,PhysRevLett.60.2089}, the quantitative impact of the
phonon dispersion is not fully understood. In particular, recent functional
renormalization group results for the model with acoustic phonons
\cite{Bakrim2015} agree quantitatively with numerical results for the model
with optical phonons \cite{PhysRevB.67.245103}. On the other hand, a
significant effect of the dispersion was reported in
Ref.~\cite{PhysRevB.83.195105}.

Here, we use quantum Monte Carlo simulations of the original SSH model (with
acoustic phonons) to demonstrate that it gives results that are identical
within statistical errors to the simplified model with optical phonons for
all parameters considered. Given this agreement, we study in detail the
real-space correlation functions and calculate spectral functions of the
simplified models. We consider the spinless and the spinful SSH model, as
well as the SSH model with additional Coulomb interaction.

The paper is organized as follows. In Sec.~\ref{sec:model} we define the
models, in Sec.~\ref{sec:method} we discuss the key methodological aspects,
in Sec.~\ref{sec:results} we present the numerical results, in
Sec.~\ref{sec:discussion} we discuss the impact of the phonon dispersion and
lattice fluctuations as well as the low-energy theory, and in
Sec.~\ref{sec:conclusions} we conclude.

\section{Models}\label{sec:model}

In addition to the original SSH model \cite{PhysRevLett.42.1698}, two other
variants have been studied in the past. First, an SSH model with optical
phonons but with the original coupling term, proposed in
Ref.~\cite{PhysRevB.56.4484}, and second a model with optical phonons and a
simplified coupling term introduced in Ref.~\cite{PhysRevB.67.245103}.

\subsection{SSH model}

Defining the bond operator
\begin{equation}
\hat{B}_{i} = \sum_\sigma\left(\hat{c}^\dag_{i\sigma}
\hat{c}^\nag_{i+1\sigma} +  \hat{c}^\dag_{i+1\sigma}
\hat{c}^\nag_{i\sigma}\right)
\end{equation}
the SSH Hamiltonian \cite{PhysRevLett.42.1698} can be written as 
\begin{eqnarray}\label{eq:SSHq} 
  \hat{H}
  = 
  -t \sum_{i} \hat{B}_{i}
  &+&g \sum_{i} \hat{B}_{i} (\oQ_{i+1}-\oQ_i)\\\nonumber
  &+&
   \sum_i
   \left[
     \frac{1}{2M} \oP^2_i  +    \frac{K}{2}  (\oQ_{i+1}-\oQ_i)^2
  \right]\,.
\end{eqnarray}
The first term describes nearest-neighbor electronic hopping with amplitude
$t$ in the undistorted chain, the second term is a coupling between electrons
and phonons in the form of a modulation of the electronic hopping as a result
of lattice distortions (see also Ref.~\cite{PhysRevLett.25.919}), and the
last term describes harmonic oscillators with mass $M$, spring constant $K$,
coordinate $\oQ_i$, and momentum $\oP_i$; it can be written as $\sum_q \om_q
\hat{b}^\dag_q \hat{b}^\nag_q$, with the phonon dispersion $\omega_q =
\om_\pi \sin (q/2)$.

The spinless SSH model has the same form as Eq.~(\ref{eq:SSHq}), with the
bond operator given by $\hat{B}_{i} = \hat{c}^\dag_{i} \hat{c}^\nag_{i+1} +
\hat{c}^\dag_{i+1} \hat{c}^\nag_{i}$ \cite{PhysRevB.27.1680}.

\subsection{Optical SSH model}

We also simulated the Hamiltonian \cite{PhysRevB.67.245103}
\begin{align}\label{eq:SSH0}
  \hat{H}
  &=
  -t \sum_{i} \hat{B}_{i}
  +
  \sum_i
  \left[
   \frac{1}{2M} \oP^2_i  +    \frac{K}{2} \oQ_i^2   
  \right]
  +
  g
  \sum_{i} 
  \hat{B}_{i}
  \oQ_{i}\,,
\end{align}
which describes the coupling of electrons to optical phonons with frequency
$\omega_0=\sqrt{K/M}$. As in the SSH model, lattice distortions modulate the
electronic hopping, but the interaction only involves the coordinate $\oQ_i$
rather than the difference $\oQ_{i+1}-\oQ_i$. In the following, we refer to
Eq.~(\ref{eq:SSH0}) as the {optical SSH   model}, and will demonstrate that
it gives the same results as the SSH model~(\ref{eq:SSHq}). The optical
phonons in Eq.~(\ref{eq:SSH0}) may be regarded as describing fluctuations
of the bonds rather than atom positions as in Eq.~(\ref{eq:SSHq}).
As discussed in Sec.~\ref{sec:discussion}, this interpretation provides
a physical understanding of the observed agreement between
Eqs.~(\ref{eq:SSHq}) and~(\ref{eq:SSH0}).

\subsection{SSH-$UV$ model}

In the spinful case, we will also study the impact of electron-electron
interactions. Following previous work
\cite{PhysRevLett.51.292,PhysRevLett.51.296,PhysRevB.57.11838,PhysRevB.67.245103,PhysRevB.83.195105},
we considered onsite and nearest-neighbor repulsion, as described by
\begin{equation}\label{eq:ee}
  \hat{H}_\text{ee} = U \sum_i (\on_{i\UP}-\oh) (\on_{i\DO}-\oh) + V \sum_i (\on_i-1) (\on_{i+1}-1)\,,
\end{equation}
with  $\on_i=\on_{i\UP}+\on_{i\DO}$. We have simulated Eq.~(\ref{eq:ee}) in
combination with the original SSH model~(\ref{eq:SSHq}) (SSH-$UV$ model) and
the simplified model~(\ref{eq:SSH0}) (optical SSH-$UV$ model).

\subsection{Spin-charge symmetry}

For our choice of zero chemical potential $\mu=0$, and on
bipartite lattices, the models~(\ref{eq:SSHq}) and~(\ref{eq:SSH0}) are invariant under the
particle-hole transformation $\hat{c}^\nag_{i(\sigma)}\mapsto
(-1)^i\hat{c}^\dagger_{i(\sigma)}$, and hence half-filled. For $U=V=0$, the spinful SSH and optical SSH models are
furthermore invariant under $\hat{c}^\nag_{i\DO}\mapsto
(-1)^i\hat{c}^\dagger_{i\DO}$ \cite{Hirsch83a}, which interchanges spin and
charge degrees of freedom. Whereas this transformation maps the repulsive to
the attractive Hubbard model \cite{PhysRevLett.66.3203}, here it implies a perfect
degeneracy of spin and charge degrees of freedom for $U=V=0$, as confirmed by
our results.

\subsection{Conventions}

We focus on half-filling, corresponding to an average
density $\las \on_i\ras=1$ for the spinful case, and to $\las \on_i\ras=0.5$ for
the spinless case. We use $t$ as the unit of energy, and set the lattice
constant and $\hbar$ to one. We define the dimensionless coupling constant
$\lambda=g^2/Kt$, and consider chains with $L$ sites and periodic
boundary conditions. The inverse temperature was chosen as $\beta t =L$.

\section{Method}\label{sec:method}

We used the continuous-time interaction-expansion (CT-INT) quantum Monte
Carlo method, which is based on a weak-coupling expansion of the partition
function, and gives exact results for finite systems \cite{Rubtsov05}.

We write the SSH model~(\ref{eq:SSHq}) in the generic form
\begin{equation}\label{eq:genfermionboson}
  \hat{H} = \hat{H}_0 + \sum_q \om_q \hat{b}^\dag_q \hat{b}^\nag_q + \gamma \sum_q (\hat{\rho}_q \hat{b}^\dag_q +
  \hat{\rho}^\dag_q \hat{b}^\nag_q)
\end{equation}
with the charge operator
\begin{align}\label{eq:chargeoperator}
  \hat{\rho}_q 
  &=
   \frac{2i}{\sqrt{L}}
     \frac{1}{\sqrt{2M\omega_q}}
   \sum_{k\sigma}  
  \hat{c}^\dag_{k\sigma} \hat{c}^\nag_{k+q\sigma} 
  \left[\sin k - \sin(k+q)\right]\,.
\end{align}

The partition function takes the form
\begin{equation}
  Z = \int D(\psib,\psi) \,e^{-S_0[\psib,\psi]} \int D(\phib,\phi) \,e^{-S_\text{ep}[\psib,\psi,\phib,\phi]}\,.
\end{equation}
Here, $S_\text{0}$ describes the noninteracting fermions, whereas the phonon
and electron-phonon contributions to Eq.~(\ref{eq:SSHq}) are contained in
$S_{\text{ep}}$. Carrying out the Gaussian integrals over the phonon degrees
of freedom \cite{Feynman55,Assaad07} gives a retarded and nonlocal
electron-electron interaction
\begin{equation}\label{eq:Saco}  
  {S}_1
  =   
  -\frac{g^2}{2K}
  \iint_0^\beta \rmd \tau \rmd \tau'
  \sum_{ij}
  B_i(\tau)
  \Gamma(i-j,\tau-\tau')
  B_j(\tau')\,,
\end{equation}
where $\Gamma(r,\tau) = L^{-1} \sum_q  e^{-i q r} D(q,\tau)$ and
\begin{equation}\label{eq:Dqtau}
  D(q,\tau) =   \frac{\omega_q}{2}\frac{\cosh[\omega_q(|\tau|-\beta/2)]}{\sinh[\omega_q\beta/2]}\,.
\end{equation}
A similar representation of the phonon-mediated interaction was given in
Ref.~\cite{PhysRevB.27.1680}; however, in the continuum approximation, the
SSH interaction becomes local in space, as characteristic of Holstein models
\cite{Ho59a}.

For the optical SSH model, the interaction reduces to
\begin{equation} \label{eq:Sopt}
  {S}_1
  =     
  -\frac{g^2}{2K}
  \iint_0^\beta \rmd \tau \rmd \tau'
  \sum_{i}
  B_i(\tau)
  D(\tau-\tau')
  B_i(\tau')\,,
\end{equation}
where $D(\tau)$ follows from Eq.~(\ref{eq:Dqtau}) by replacing
$\om_q\mapsto\omega_0$. Although both bond operators in Eq.~(\ref{eq:Sopt})
carry the same index $i$, the interaction remains nonlocal because
$B_i(\tau)$ involves the sites $i$ and $i+1$.

The interactions described by Eqs.~(\ref{eq:Saco}) and~(\ref{eq:Sopt}) can be
simulated with the CT-INT method by implementing a general vertex with four
lattice indices, two time indices, and two spin indices. For the optical SSH
models, there is no sign problem so that a sampling over additional Ising
spins \cite{Assaad07} is not necessary. 

In the case of the SSH-$UV$ model, the additional Coulomb interaction
described by Eq.~(\ref{eq:ee}) enters the partition function in the form of
\begin{align}
  S_2
  =
  U&
  \int_0^\beta \rmd \tau
  \sum_i
  [n_{i\UP}(\tau)-\oh] [n_{i\DO}(\tau)-\oh]
  \\\nonumber  
  +&V
  \int_0^\beta \rmd \tau
  \sum_i\sum_{\sigma\sigma'}
  [n_{i\sigma}(\tau)-\oh] [n_{i+1\sigma'}(\tau)-\oh]\,.
\end{align}
The implementation of such interactions within the CT-INT method, including
the use of auxiliary Ising fields to avoid a sign problem, has been
discussed in Refs.~\cite{Assaad07,PhysRevB.85.195115}. The
simultaneous sampling of the interactions $S_1$ and $S_2$ is done by randomly
choosing a vertex type.

The numerical effort scales as $(\alpha \beta L)^3$ with $\alpha\sim\lambda$ for $U=V=0$, and
$\alpha\sim U$ for $U\gg V,\lambda$.

\subsection{Sign problem}

Previous applications of the CT-INT method to electron-phonon models only
considered the case of optical phonons. For the SSH model, the momentum
dependence of the phonon dispersion $\omega_q$ leads to a pronounced minus
sign problem whose origin can be traced back to the fact that
$\Gamma(r,\tau)$ (and hence the configuration weight) can take on negative
values for $r>0$ in the case of dispersive phonons. The sign problem can be
improved to some extent (at the expense of a higher expansion order) by
generalized offsets $\alpha(\tau,s)$ (depending on Ising spins $s=\pm1$,
cf. Ref.~\cite{Assaad07}) in the bond operators $B_i(\tau)$ for vertices with
negative weight.

\subsection{Observables}

We calculated the real-space correlation functions 
\begin{align}
  C_\text{b}(r) & = \las (\hat{B}_{r}-b)(\hat{B}_{0}-b)\ras && (\text{bond})\\
  C_{\rho}(r) &=  \las (\on_r -n) (\on_0-n) \ras && (\text{charge})\\
  C_{\sigma}(r)  &= \las \hat{s}^x_r \hat{s}^x_0 \ras && (\text{spin})\\
  C_\text{p}(r) &= \las \hat{\Delta}^\dag_r \hat{\Delta}^\nag_0 \ras &&  (\text{pairing})
\end{align}

Here, $n=\las\on_i\ras$, $b=\las \hat{B}_i\ras$, and we consider the
transverse spin correlations ($\las \hat{s}^x_r \hat{s}^x_0 \ras$ is identical
within error bars to $\las \hat{s}^z_r \hat{s}^z_0 \ras$, as required by
symmetry). The pairing operator $\hat{\Delta}_r$ is defined as $\hat{\Delta}_r =
c^\dag_{r\UP} c^\dag_{r\DO}$ for spinful fermions (singlet pairing), and as
$\hat{\Delta}_r = c^\dag_{r\vphantom{+1}} c^\dag_{r+1}$ for spinless fermions (extended
singlet pairing).

In combination with the maximum entropy method \cite{Beach04a}, we 
calculated the single-particle spectral function 
\begin{align}\label{eq:akw}
  A(k,\om)
  =
  \frac{1}{Z}\sum_{ij}
  &
  {|\bra{i} \hat{O}_k \ket{j}|}^2 
  (e^{-\beta E_i}+e^{-\beta E_j})
  \delta(\Delta_{ji}-\om)
  \,,
\end{align}
with $\hat{O}_k=c_k$ for spinless fermions, and $\hat{O}_k=c_{k\sigma}$ for
spinful fermions. Additionally, we calculated 
\begin{align}\label{eq:nqw}
  S_\alpha(q,\om)
  &=
  \frac{1}{Z}\sum_{ij} {|\bra{i} \hat{O}^\alpha_q \ket{j}|}^2
  e^{-\beta E_i} 
  \delta(\Delta_{ji}-\om)
  \,,
\end{align}
with $\hat{O}^\alpha_q$ either a bond, charge, or spin operator. Here, 
$\ket{i}$ is an eigenstate with energy $E_i$, and $\Delta_{ji}=E_j-E_i$.

\section{Results}\label{sec:results}

\subsection{SSH model vs. optical SSH model}\label{sec:results:comparison}

Several different SSH-type models have been studied in the literature.  The
phase diagram for a half-filled band has been investigated numerically in
terms of the original Hamiltonian~(\ref{eq:SSHq}) in
Ref.~\cite{PhysRevB.27.1680} (with a cutoff to avoid an unphysical sign
change of the hopping term for large distortions), in
Ref.~\cite{PhysRevB.73.045106} by solving the original SSH model, in
Ref.~\cite{PhysRevB.67.245103} by considering the optical SSH
model~(\ref{eq:SSH0}), and in Ref.~\cite{PhysRevB.83.195105} by simulating
the original interaction term in combination with a general phonon dispersion
and extrapolating to acoustic and optical phonons \cite{PhysRevB.83.195105}.

\begin{figure}[t]
  \includegraphics[width=0.45\textwidth]{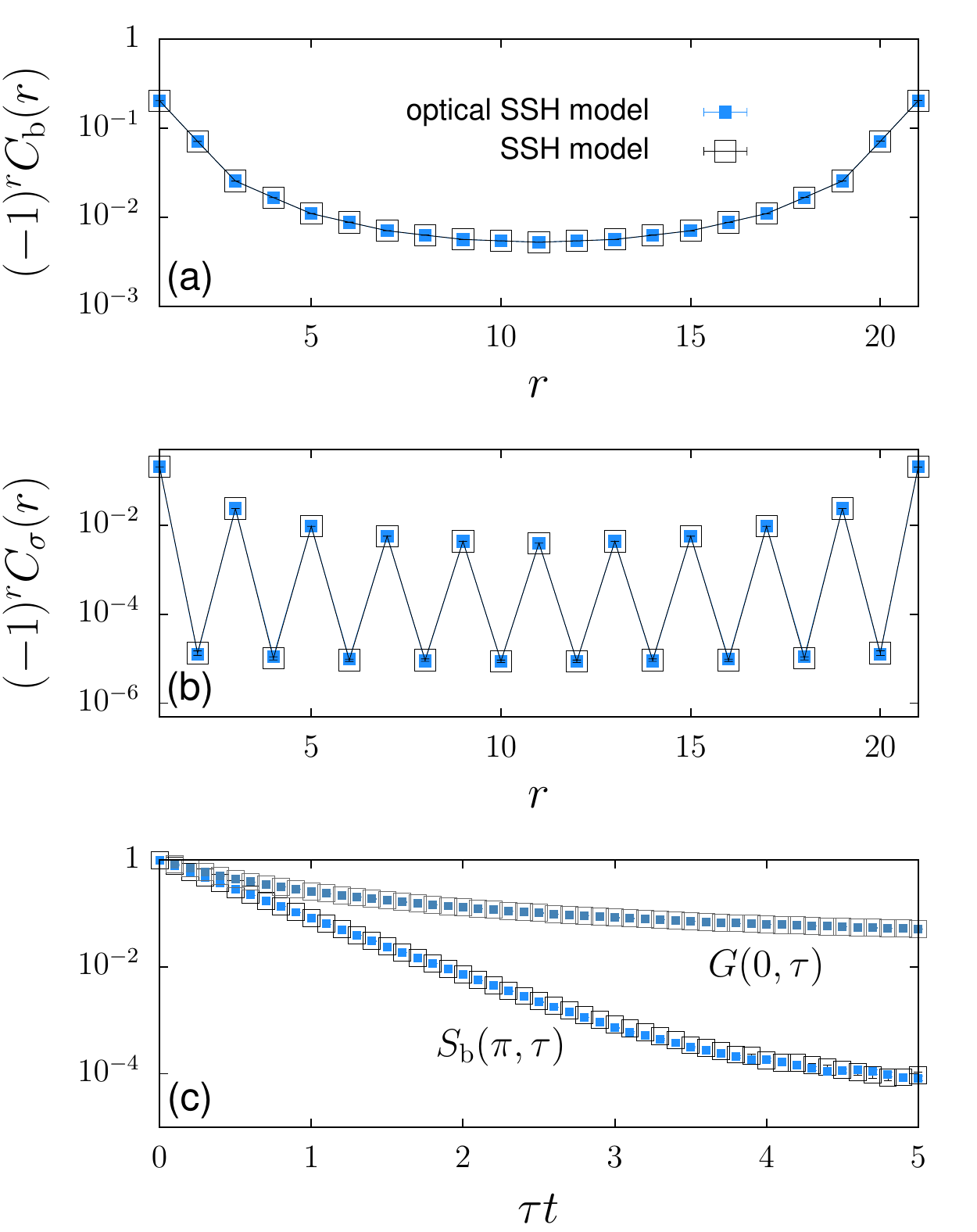}  
  \caption{\label{fig:comparison} (Color online) Comparison between the
    optical SSH model (filled symbols) and the SSH model (open symbols) for
    (a) real-space bond correlations, (b) real-space spin correlations, (c)
    the single-particle Green function $G(k=0,\tau)$ and the dynamic bond
    correlation function $S_\text{b}(q=\pi,\tau)$.  Here, $\beta t =L=22$,
    $\om_0=\om_\pi=0.1t$, $U=V=0$, and $\lambda=0.2$.}
\end{figure}

Here, we present a comparison of the SSH model~(\ref{eq:SSHq}) and the
optical SSH model~(\ref{eq:SSH0}), taking $\omega_0=\omega_\pi=0.1t$ to
alleviate the sign problem; a larger value $\omega_0=\omega_\pi=0.5t$ was
used for most other results.  Figure~\ref{fig:comparison}(a) shows the bond
correlation function $C_\text{b}(r)$, whereas Fig.~\ref{fig:comparison}(b)
shows the corresponding spin correlations. Remarkably, the results for the
models~(\ref{eq:SSHq}) and~(\ref{eq:SSH0}) are identical within statistical
errors. This quantitative agreement also extends to dynamic correlation
functions, as demonstrated for the single-particle Green function at momentum
$k=0$ and the dynamic bond correlation function at $q=\pi$ in
Fig.~\ref{fig:comparison}(c).  In addition to the results shown in
Fig.~\ref{fig:comparison}, we compared bond, charge, spin, and pairing
correlation functions, as well as the momentum distribution function, for
several other parameter sets, including nonzero $U$ and $V$, higher
temperatures, stronger coupling, and spinless models. All results were
identical within error bars to those for the original SSH coupling and
acoustic phonons. Based on this agreement, whose origin will be discussed in
Sec.~\ref{sec:discussion}, we avoided the sign problem by simulating the
corresponding sign-free optical SSH models.

\subsection{Spinless fermions}\label{sec:results:spinless}

\begin{figure}[t]
  \includegraphics[width=0.45\textwidth]{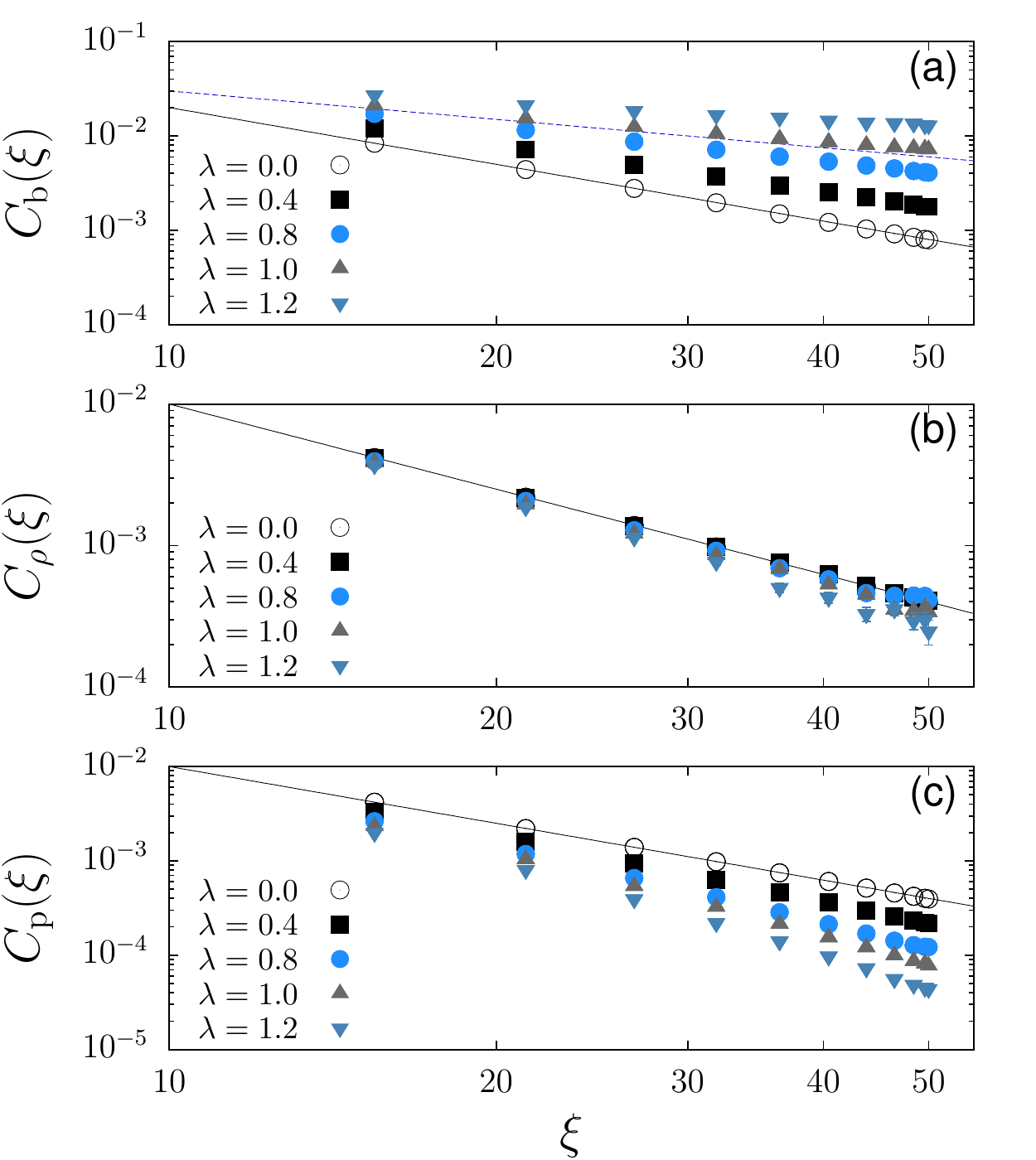}  
  \caption{\label{fig:correlations_spinless} (Color online) (a) Bond, (b)
    charge, and (c) pairing correlations of the spinless optical SSH model as
    a function of $\xi= L\sin \left(\pi r/L\right)$ (see text).  Here, $\beta
    t =L=50$, and $\om_0=0.5t$. The solid lines correspond to $1/r^2$, the
    dashed line corresponds to $1/r$. Here and in
    Figs.~\ref{fig:correlations_spinful}
    and~\ref{fig:correlations_UV_spinful}, the correlation functions are
    plotted only for odd distances $r$.}
\end{figure}

Given the theoretical prediction of an extended metallic phase without
long-range bond order in the spinless SSH model \cite{PhysRevB.27.1680}, we show in
Fig.~\ref{fig:correlations_spinless} the real-space
bond, charge, and pairing correlation functions defined in
Sec.~\ref{sec:method}. To eliminate boundary effects, we use the conformal
distance $\xi= L\sin \left(\pi r/L\right)$ \cite{Cardybook}. This allows us
to directly compare to the bosonization results for a spinless Luttinger liquid
 \cite{Voit94,Giamarchi}
\begin{align}\label{eq:correl:LLspinless}\nonumber
   C_\text{b}(r) 
  &=
  \frac{A_\text{b}}{r^{2\Kr}}\cos(2\kF r)\,,
  \\\nonumber
   C_\rho(r) 
  &=
  -
  \frac{\Kr}{2\pi^2r^2} + \frac{A_\rho}{r^{2\Kr}}\cos(2\kF r)\,,
  \\
  C_\text{p}(r) 
  &=
  \frac{A_\text{p}}{r^{2\Kr^{-1}}} \,.
\end{align}

For $\lambda=0$, all correlators decay with a power law $1/r^2$. According to
our results, bond correlations become dominant for $\lambda>0$, suggesting
$K_\rho<1$ and hence a repulsive Luttinger liquid, in accordance with
Ref.~\cite{Ba.Bo.07}. At the same time, charge and pairing correlations are
suppressed. While the decay of pairing correlations can be explained by the
increase of the exponent $2K_\rho^{-1}$, $2\kF$ charge correlations formally
have the same exponent as $2\kF$ bond correlations, see
Eq.~(\ref{eq:correl:LLspinless}).  The absence of enhanced charge
correlations in Fig.~\ref{fig:correlations_spinless}(b) can be
attributed to a suppression of the amplitude, $A_\rho\to0$
\cite{PhysRevB.45.4027}, causing the first term ($1/r^2$) to
dominate in the metallic phase [the exponent is close to $2$ in
Fig.~\ref{fig:correlations_spinless}(b)]. Pairing correlations
[Fig.~\ref{fig:correlations_spinless}(c)] are always subdominant in the
metallic phase. In the insulating phase, charge and pairing correlations are
strongly suppressed---consistent with an exponential decay---while bond
correlations become long-ranged.

\begin{figure}[t]
  \includegraphics[width=0.45\textwidth]{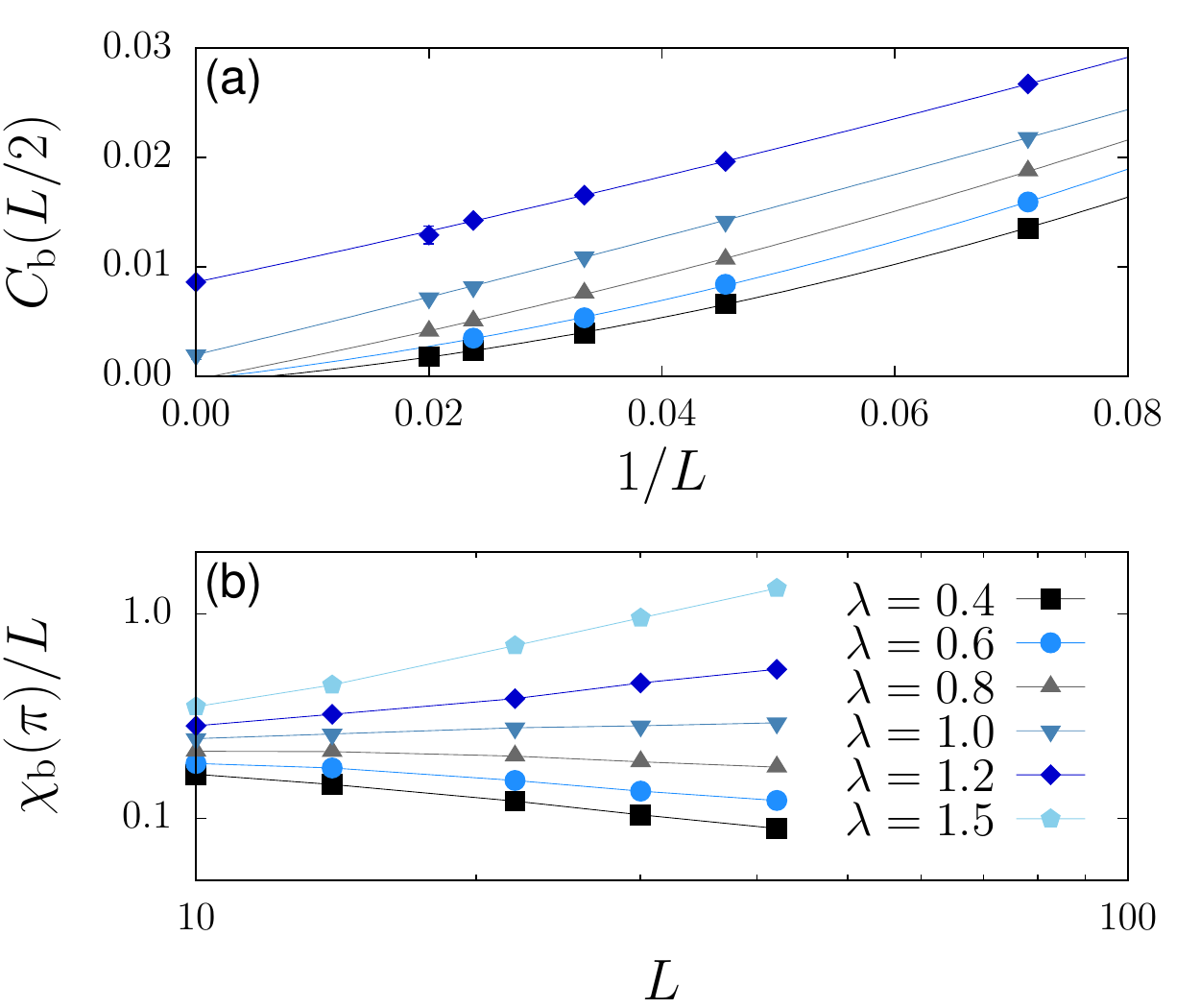}  
  \caption{\label{fig:longrangeorder-spinless} (Color online) (a) Finite-size
    scaling of the bond correlations at the largest distance $r=L/2$, using a
    second-order polynomial extrapolation.  (b) Finite-size scaling of the
    bond susceptibility for the spinless optical SSH model.  Here, $\beta t
    =L$, and $\om_0=0.5t$. The key in (b) applies to both panels.}
\end{figure}

To determine the critical value $\lambda_{c}$ for the Peierls
transition, we show in Fig.~\ref{fig:longrangeorder-spinless}(a) a
finite-size scaling of the bond correlations at the largest distance
$r=L/2$. The extrapolation using system sizes up to $L=50$ suggests
$\lambda_{c}=0.9(1)$. The onset of long-range order can also be tracked
by calculating the bond susceptibility \cite{PhysRevB.67.245103}
\begin{equation}\label{eq:chic}
  \chi_\text{b}(\pi) = \frac{1}{L} \sum_{ij} (-1)^{i-j} \int_0^\beta d\tau
  \las \hat{B}_i(\tau) \hat{B}_j(0) \ras\,.
\end{equation}
As a function of $L$ (and for $\beta=L$), $\chi_\text{b}(\pi)/L$ is expected
to vanish in the metallic phase, and to diverge $\sim L$ in the dimerized
Peierls phase. For spinless fermions, $\chi_\text{b}$ has a more favorable
scaling with system size than the bond correlations shown in
Fig.~\ref{fig:longrangeorder-spinless}(a), and does not involve an unknown
scaling function. Our numerical results in
Fig.~\ref{fig:longrangeorder-spinless}(b) confirm the expected behavior of
the susceptibility, and give a critical value consistent with
Fig.~\ref{fig:longrangeorder-spinless}(a).  We also carried out simulations
for other phonon frequencies, and found that $\lambda_c$ increases with
increasing $\omega_0$, in agreement with theoretical expectations
\cite{PhysRevB.27.1680}.

Within a bosonization description of the Luttinger liquid to
(charge-density-wave) insulator transition, $K_\rho=1/2$ corresponds to the
critical point of the Kosterlitz-Thouless phase
transition. Equation~(\ref{eq:correl:LLspinless}) therefore suggests a $1/r$
decay at the critical point. Although we have to consider the possibility of
logarithmic corrections in the vicinity of the transition, the correlation
functions in Fig.~\ref{fig:correlations_spinless}(a) indeed decay slightly
faster than $1/r$ (consistent with $K_\rho>1/2$) for $\lambda=0.8$, and
slightly slower than $1/r$ for $\lambda=1$, consistent with
$\lambda_c=0.9(1)$.

\begin{figure}[t]
  \includegraphics[width=0.45\textwidth]{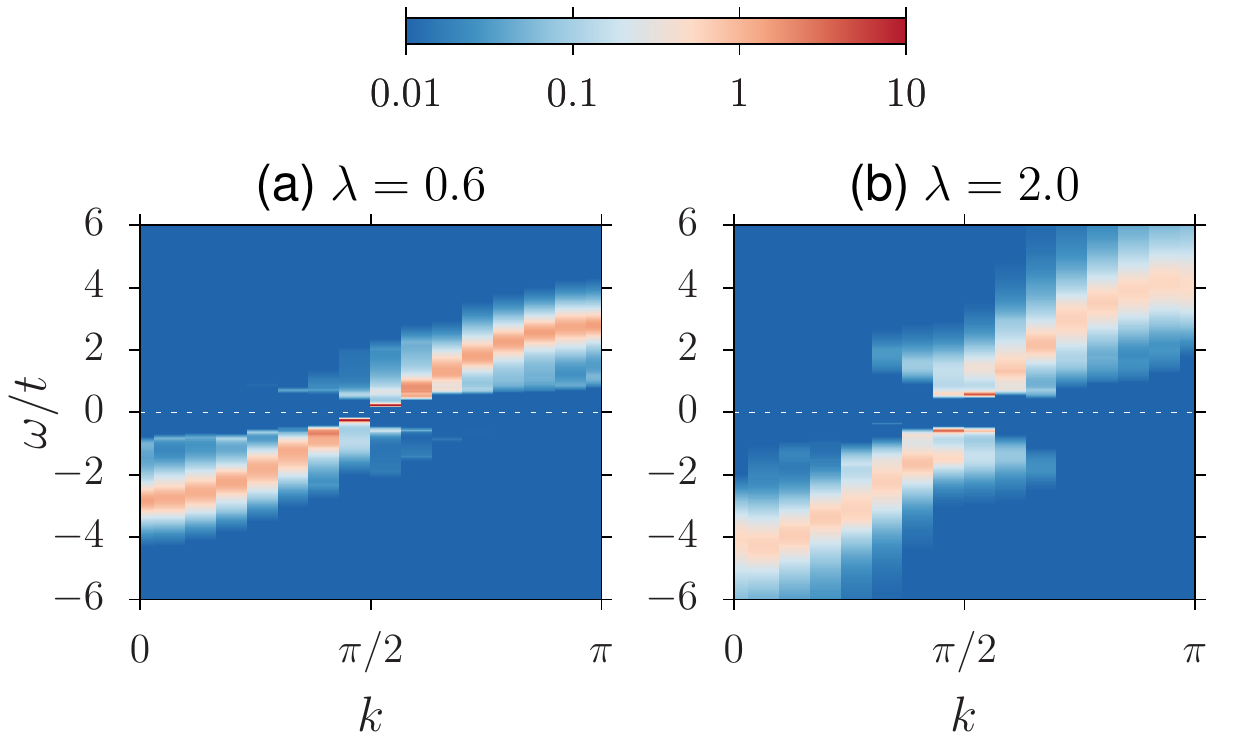}  
  \caption{\label{fig:dynamics_spinless_akw} (Color online) Single-particle
    spectral function $A(k,\omega)$ of the spinless optical SSH model. The dashed line indicates
    the Fermi level. Here, $\beta t =L=30$,
    $\om_0=0.5t$. Color scheme for this and other density plots from Ref.~\cite{colorscheme}.
}
\end{figure}

To further characterize the two phases, we calculated spectral
functions. Figure~\ref{fig:dynamics_spinless_akw} shows the single-particle
spectral function for (a) $\lambda=0.6$ (metallic phase) and (b) $\lambda=2$
(insulating phase). In Fig.~\ref{fig:dynamics_spinless_akw}(a) we see a
gapless (within the available momentum resolution) band derived from the bare
dispersion $-2t\cos k$.  The most prominent effects of the electron-phonon
coupling are (i) a substantial broadening outside the coherent interval
$[-\om_0,\om_0]$, and (ii) an enhanced bandwidth [$\approx 6t$ in Fig.~\ref{fig:dynamics_spinless_akw}(a)]
arising from the enhanced hopping due to the SSH interaction, see Eq.~(\ref{eq:SSH0}). (The bandwidth of the
spectrum remains $\approx 4t$ in Holstein-type models \cite{Hohenadler10a}.) In the Peierls
phase, Fig.~\ref{fig:dynamics_spinless_akw}(b), the spectrum has acquired a
gap at the Fermi level, and exhibits signatures of backfolded shadow bands
characteristic of systems with reduced periodicity
\cite{Vo.Pe.Zw.Be.Ma.Gr.Ho.Gr.00}. While these features are qualitatively
captured by mean-field theory \cite{Hohenadler10a}, at and near $\kF$, we can
identify two separate energy scales in
Fig.~\ref{fig:dynamics_spinless_akw}(b) previously discussed for the spinless
Holstein model \cite{Hohenadler10a}. The
lowest-energy excitation can be related to soliton excitations
\cite{Hohenadler10a}, whereas the high-energy excitations correspond to the
renormalized cosine band structure with a dimerization gap.

\begin{figure}[t]
  \includegraphics[width=0.45\textwidth]{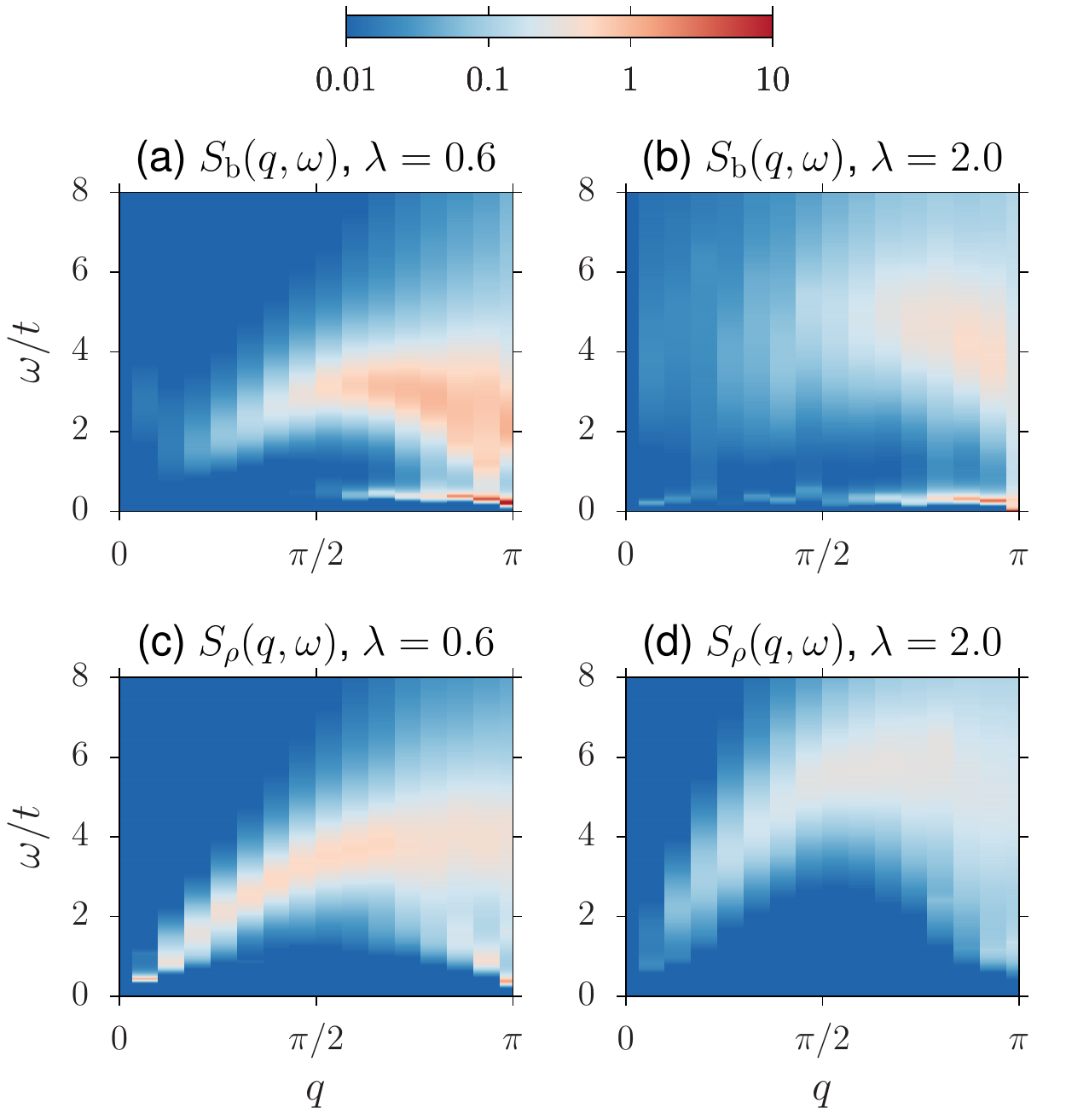}  
  \caption{\label{fig:dynamics_spinless_ph} (Color online) (a)--(b) Dynamic bond
    and (c)--(d) dynamic density structure factor of the spinless optical
    SSH model. Here, $\beta t =L=30$, and $\om_0=0.5t$.}
\end{figure}

Figure~\ref{fig:dynamics_spinless_ph} shows the corresponding dynamic bond
and density structure factors. For weak coupling, the latter still resemble
the noninteracting particle-hole continuum. Additionally, the bond structure factor in
Figs.~\ref{fig:dynamics_spinless_ph}(a) and (b) reveals at low
energies the renormalized phonon frequency, in particular a central ($\om=0$)
peak at $q=\pi$ in the Peierls phase in Fig.~\ref{fig:dynamics_spinless_ph}(b) related
to long-range order (in the Holstein model, where the lattice is coupled to
the charge, the phonon dispersion appears in the dynamic charge structure
factor \cite{Hohenadler10a,WeAsHo15I}). The density structure factor in the metallic phase,
Fig.~\ref{fig:dynamics_spinless_ph}(c), features a gapless linear mode at
small $q$, whereas a gap and a strong suppression of low-energy excitations
are visible in the Peierls phase, see Fig.~\ref{fig:dynamics_spinless_ph}(d).

\subsection{Spinful fermions}\label{sec:results:spinful}

\begin{figure}[t]
  \includegraphics[width=0.45\textwidth]{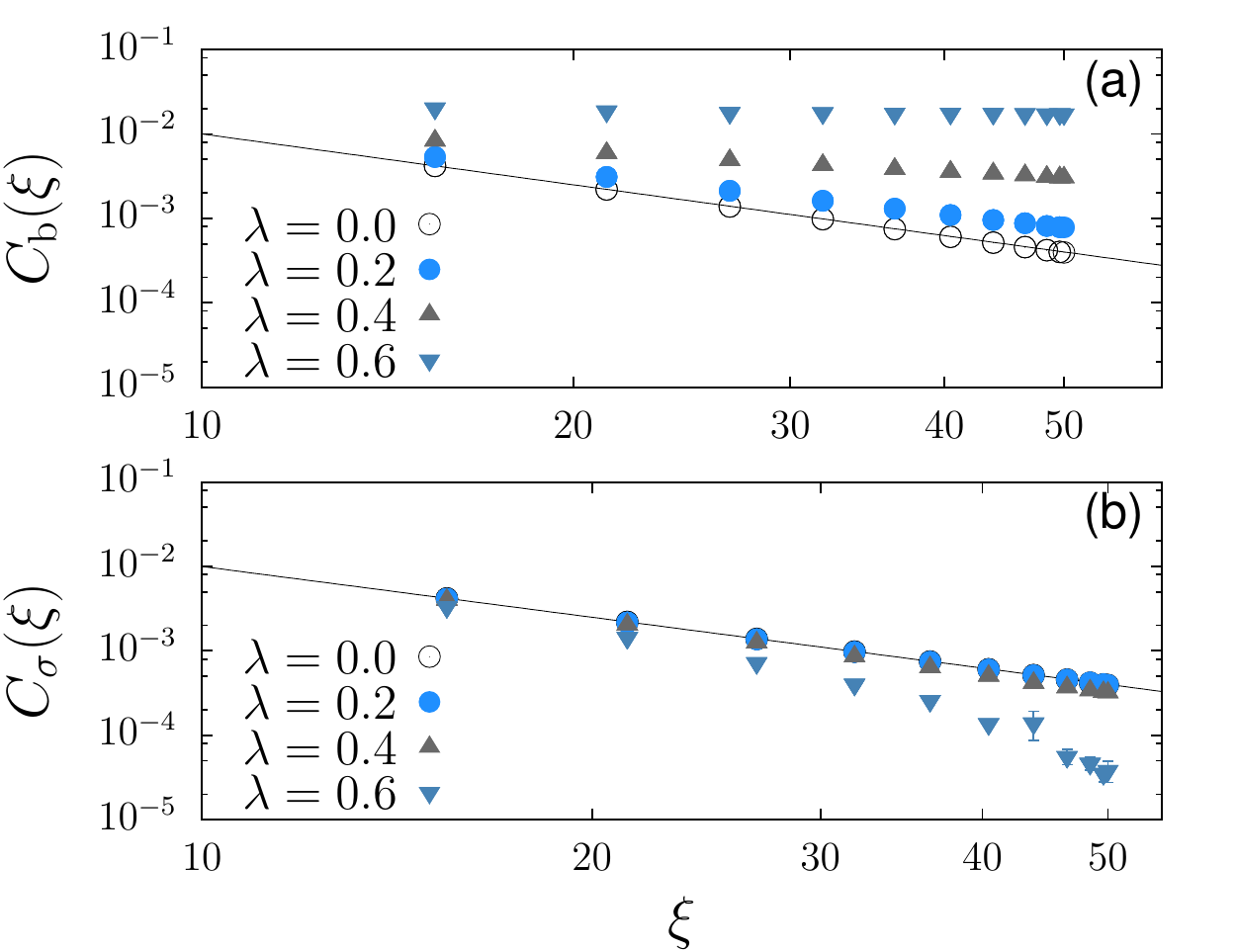}  
  \caption{\label{fig:correlations_spinful} (Color online) (a) Bond and (b)
    spin correlations of the spinful optical SSH model as a function of  $\xi= L\sin \left(\pi r/L\right)$.
    Here, $\beta t =L=50$, and $\om_0=0.5t$.  The solid lines correspond to $1/r^2$}
\end{figure}

\begin{figure}[t]
  \includegraphics[width=0.45\textwidth]{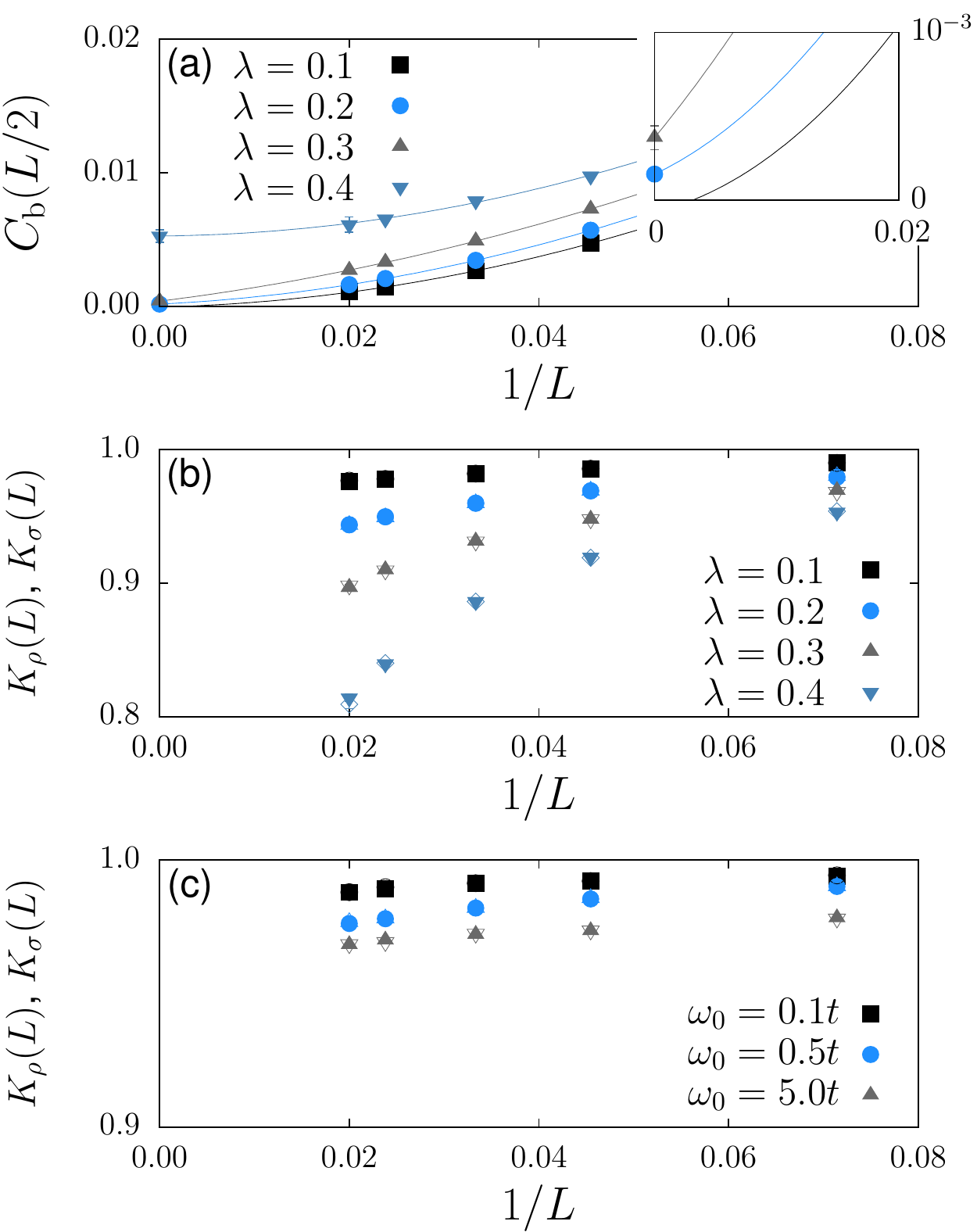}  
  \caption{\label{fig:longrangeorder_spinful} (Color online) (a) Finite-size scaling of
    the bond correlations at the largest distance $r=L/2$ for the spinful optical
    SSH model and different values of $\lambda$. Here, $\beta t=L$, and
    $\om_0=0.5t$.  The  inset shows a closeup of the extrapolated values. (b)
    Finite-size estimates of the Luttinger liquid
    parameters [we find $K_\rho(L)=K_\sigma(L)$ within error bars] for the
    same parameters as in (a). (c) 
   $K_\rho(L)$ and $K_\sigma(L)$ for different phonon
    frequencies. Here, $\beta t =L$, and $\lambda=0.1$.}
\end{figure}

We begin by considering the real-space correlation functions shown in
Fig.~\ref{fig:correlations_spinful}. Similar to the spinless case, bond
correlations decay increasingly slowly upon switching on the electron-phonon
coupling. At the same time, we observe a suppression of charge, pairing and
spin correlations [spin correlations are shown in
Fig.~\ref{fig:correlations_spinful}(b), the others behave similarly]. Since
we expect an ordered Peierls state for any $\lambda>0$, bond correlations
should oscillate with a constant, nonzero amplitude at large distances, and
all other correlation functions should decay exponentially. While the
finite-size extrapolation of the bond correlations at the largest distance in
Fig.~\ref{fig:longrangeorder_spinful}(a) gives a nonzero value already for
$\lambda=0.2$, the expected exponential decay of spin correlations is only
clearly visible (for the system size used) for $\lambda=0.6$ in
Fig.~\ref{fig:correlations_spinful}(b).

The Peierls state can be detected even at weak coupling. Because the bond
susceptibility is not a reliable indicator to track the onset of long-range
order in systems with a spin gap \footnote{%
A divergent susceptibility is a necessary but not a sufficient condition for
the existence of long-range order. In particular, a divergence occurs if the
bond correlator in Eq.~(\ref{eq:chic}) decays with an exponent less or equal
to one. Such a decay is characteristic of repulsive Luther-Emery liquids with
a spin gap but without long-range order.}, we instead consider the
finite-size estimates of the Luttinger liquid parameters $K_\rho$ and
$K_\sigma$. The latter are defined in terms of the charge and spin structure
factors as
\begin{align}\label{eq:krhoL}\nonumber
K_\rho(L) &= \pi  S_\rho(q_1) / q_1\,,\\
K_\sigma(L) &= \pi  S_\sigma(q_1) / q_1\,,
\end{align}
where $q_1=2\pi/L$ is the smallest nonzero wave vector for system size
$L$. Finite-size extrapolation has been shown to give accurate results for
the Luttinger parameters, for example, in the extended Hubbard model
\cite{PhysRevB.74.245110}. However, as
argued in Refs.~\cite{PhysRevB.65.155113,PhysRevB.67.245103}, even very small
spin gaps can be detected from values $K_\sigma(L)<1$, as they imply $K_\sigma(L\to\infty)=0$. In
contrast, in systems with gapless spin excitations, numerical simulations
usually give $K_\sigma(L)>1$ with a very
slow (logarithmic) convergence to the value $K_\sigma=1$ implied by the
$SU(2)$ spin symmetry. This diagnostic gives reliable results for the
integrable attractive and repulsive Hubbard models \cite{hardikar:245103}. 
For the SSH model, we find $K_\sigma(L)<1$ and hence a spin gap even for
$\lambda=0.1$ [Fig.~\ref{fig:longrangeorder_spinful}(b)], which is consistent
with a Peierls state for any $\lambda>0$. The same behavior is observed over
the whole range of phonon frequencies from the adiabatic to the antiadiabatic
regime, see Fig.~\ref{fig:longrangeorder_spinful}(c).

In general, a spin gap does not necessarily imply long-range Peierls
order. For example in the Holstein model, an extended metallic (Luther-Emery
\cite{Lu.Em.74}) phase with a spin gap but gapless charge excitations
\cite{PhysRevB87.075149} exists. While theory suggests that the spin and
charge gaps open simultaneously in the SSH model, we can obtain explicit
evidence from the degeneracy of $K_\rho(L)$ and $K_\sigma(L)$ in
Fig.~\ref{fig:longrangeorder_spinful}(b).  Within statistical errors, the
correlation functions $S_\rho(q,\tau)$ and $S_\sigma(q,\tau)$ and hence also
$K_\rho(L)$ and $K_\sigma(L)$ are identical. This is a consequence of the
spin-charge symmetry discussed in Sec.~\ref{sec:model}, and has important
consequences: if $K_\sigma(L)<1$ indicates a spin gap,
$S_\rho(q,\tau)=S_\sigma(q,\tau)$ implies also a charge gap and hence
insulating behavior. Figure~\ref{fig:longrangeorder_spinful}(b) therefore
suggests that the spinful optical SSH model is a Peierls insulator even for
very small $\lambda$, in agreement with the theoretical prediction of
$\lambda_c=0$. Given the quantitative agreement between the optical SSH model
and the SSH model demonstrated in Fig.~\ref{fig:comparison}, our findings
carry over to Eq.~(\ref{eq:SSHq}).

\begin{figure}[t]
  \includegraphics[width=0.45\textwidth]{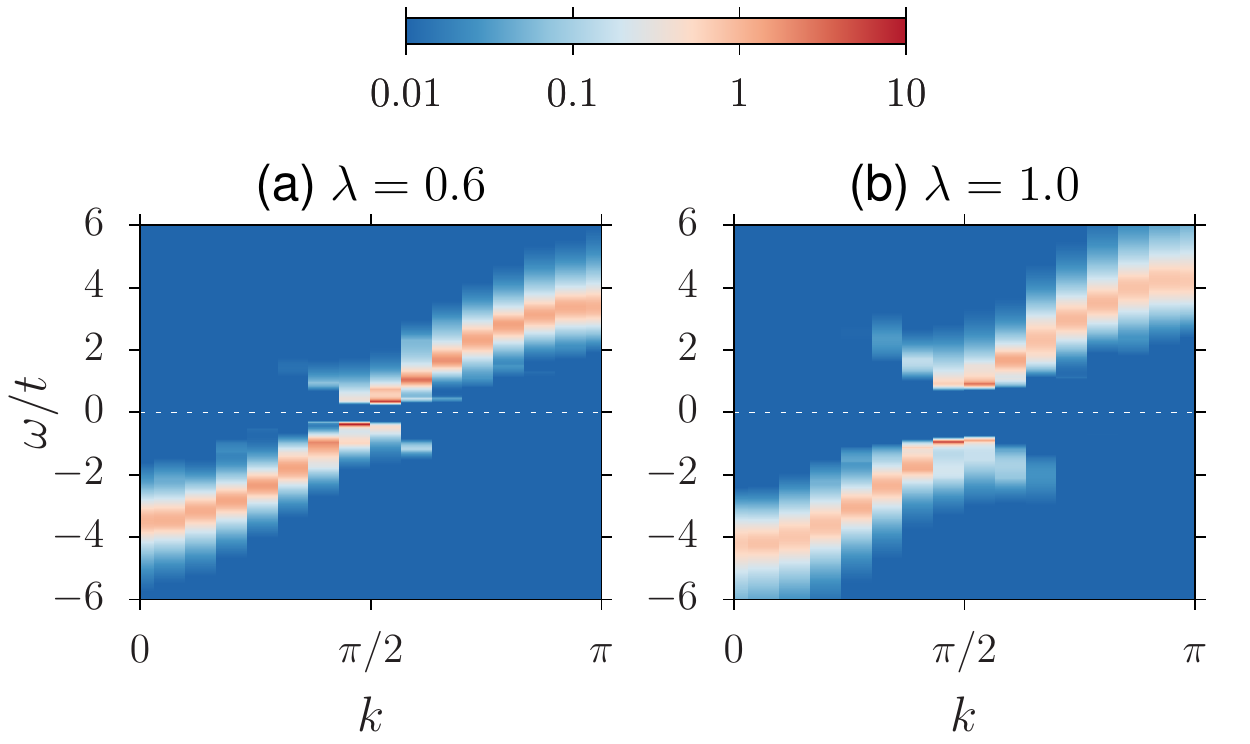}  
  \caption{\label{fig:dynamics_spinful_akw} (Color online) Single-particle
    spectral function $A(k,\omega)$ of the spinful optical SSH model. The dashed line indicates
    the Fermi level. Here, $\beta t =L=30$, and $\om_0=0.5t$.}
\end{figure}

Having established the insulating behavior for any $\lambda>0$, we now
consider excitation spectra. Figure~\ref{fig:dynamics_spinful_akw} shows the
single-particle spectral function for $\lambda=0.6$ and $\lambda=1$.  The gap
opens exponentially starting from $\lambda=0$. Nevertheless, we see a small
gap and backfolded shadow bands in Fig.~\ref{fig:dynamics_spinful_akw}(a), in
contrast to the results for the metallic phase of the spinless model in
Fig.~\ref{fig:dynamics_spinless_akw}(a) for the same value of the coupling
($\lambda=0.6$).  For a stronger coupling $\lambda=1$,
Fig.~\ref{fig:dynamics_spinful_akw}(b) reveals a substantial gap, and the
spectrum closely resembles that of the spinless SSH model in the Peierls
phase, see Fig.~\ref{fig:dynamics_spinless_akw}(b).

Figure~\ref{fig:dynamics_spinful_lambda0.6} shows the dynamic bond and spin
structure factors for $\lambda=1$. As expected for the dimerized phase,
Fig.~\ref{fig:dynamics_spinful_lambda0.6}(a) features a central peak with
large spectral weight at $q=\pi$. Both charge and spin excitations (identical
within error bars) are gapped at small $q$.

\subsection{Spinful fermions with Coulomb interaction}\label{sec:results:UV}

We now turn to the spinful SSH-$UV$ model with $U=4V=2.5t$, considered to be
appropriate for polyacetylene \cite{PhysRevB.57.11838}, and a phonon
frequency $\om_0=0.5t$. In principle, our method can also be applied to
models with long-range interactions \cite{Ho.As.Fe.12}. Because
Eqs.~(\ref{eq:SSHq}) and~(\ref{eq:SSH0}) give identical results even with the
additional electron-electron interaction, we simulated the optical SSH model
to avoid the minus-sign problem mentioned in Sec.~\ref{sec:method}.

From Refs.~\cite{PhysRevB.29.4230,PhysRevB.67.245103,PhysRevB.83.195105}, it
is known that the phase diagram consists of Mott and Peierls insulating
regions, except for the adiabatic limit where the Peierls state seems to
remain stable for any $\lambda>0$.  Away from this limit, the system is a
Mott insulator for $\lambda<\lambda_c$, with gapless spin excitations but a
gap for single-particle and charge excitations. For $\lambda=0$ (extended
Hubbard model), the ground state is a Mott insulator for $U\gg2V$
\cite{PhysRevB.45.4027,PhysRevB.65.155113,PhysRevLett.99.216403}, as relevant
for polyacetylene. The electron-phonon coupling competes with the dominant
antiferromagnetic spin correlations, and eventually gives rise to a Peierls
phase with a spin gap and long-range bond order
\cite{PhysRevB.67.245103,PhysRevB.83.195105}.

\begin{figure}[t]
  \includegraphics[width=0.45\textwidth]{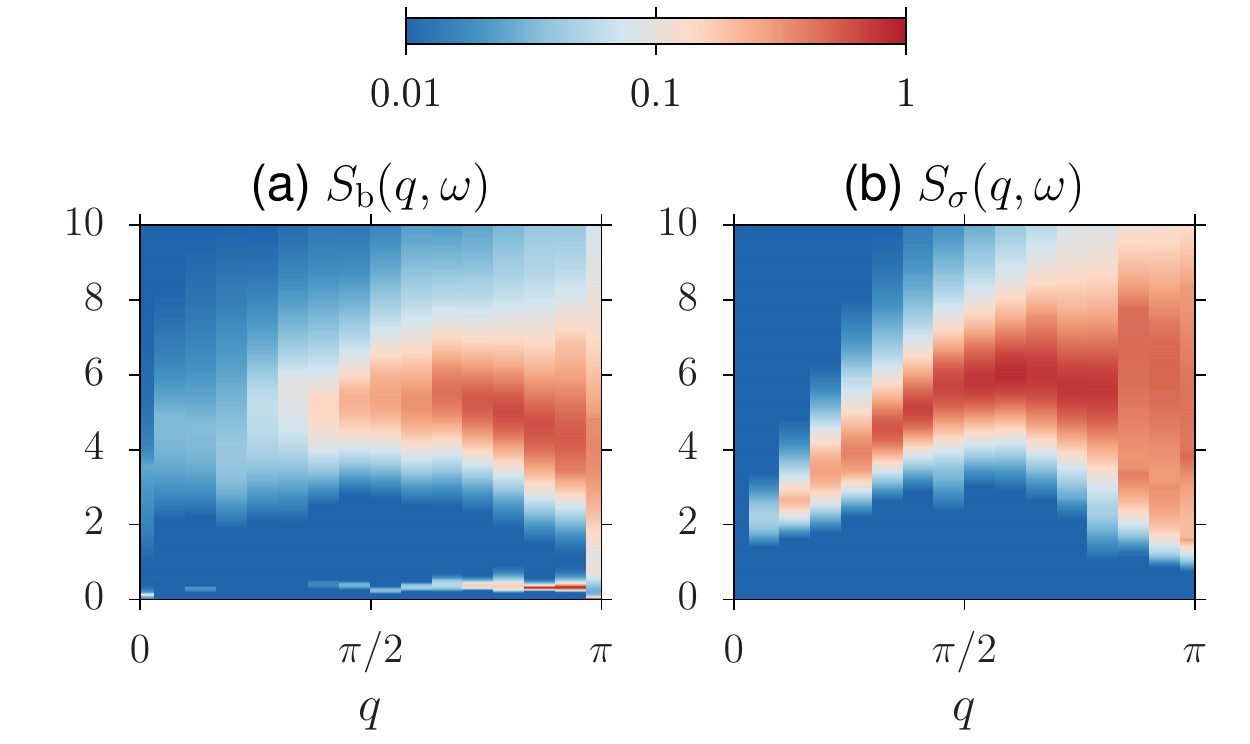}  
  \caption{\label{fig:dynamics_spinful_lambda0.6} (Color online) (a) Dynamic
    bond and (b) dynamic spin structure factor of 
    the spinful optical SSH model.  Here, $\beta t =L=30$, $\om_0=0.5t$, and
    $\lambda=1$.}
\end{figure}

\begin{figure}[t]
  \includegraphics[width=0.45\textwidth]{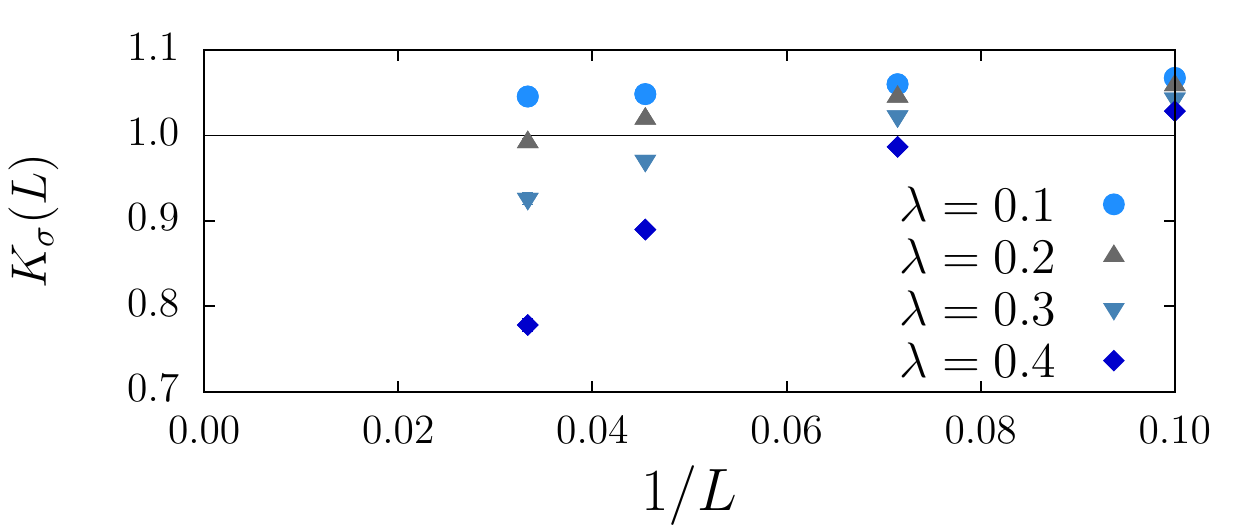}  
  \caption{\label{fig:longrangeorder_UV_spinful} (Color online) Finite-size scaling of
    $K_\sigma(L)$ for the optical SSH-$UV$ model.
    Here, $\beta t =L$, $\om_0=0.5t$, $U=2.5t$, and
    $V=0.625t$.}
\end{figure}

To estimate $\lambda_c$ for the Mott-Peierls transition for our choice of
parameters, we carried out a finite-size scaling of $K_\sigma(L)$; the
additional Coulomb interaction restricts us to system sizes $L\leq 30$. The
results in Fig.~\ref{fig:longrangeorder_UV_spinful} for $\lambda=0.1$ are
consistent with the absence of a spin gap and $K_\sigma=1$, and therefore
with a Mott insulating state. In contrast, for $\lambda\geq0.2$, we have
$K_\sigma(L)<1$ and a finite-size dependence consistent with $K_\sigma=0$ and
a spin-gapped Peierls phase.  Moreover, we find $K_\rho(L)<1$ for all values
of $\lambda$ considered, and a very weak dependence of $K_\rho(L)$ on
$\lambda$ in the range $0\leq\lambda\leq 0.4$ (not shown).

\begin{figure}[t]
  \includegraphics[width=0.45\textwidth]{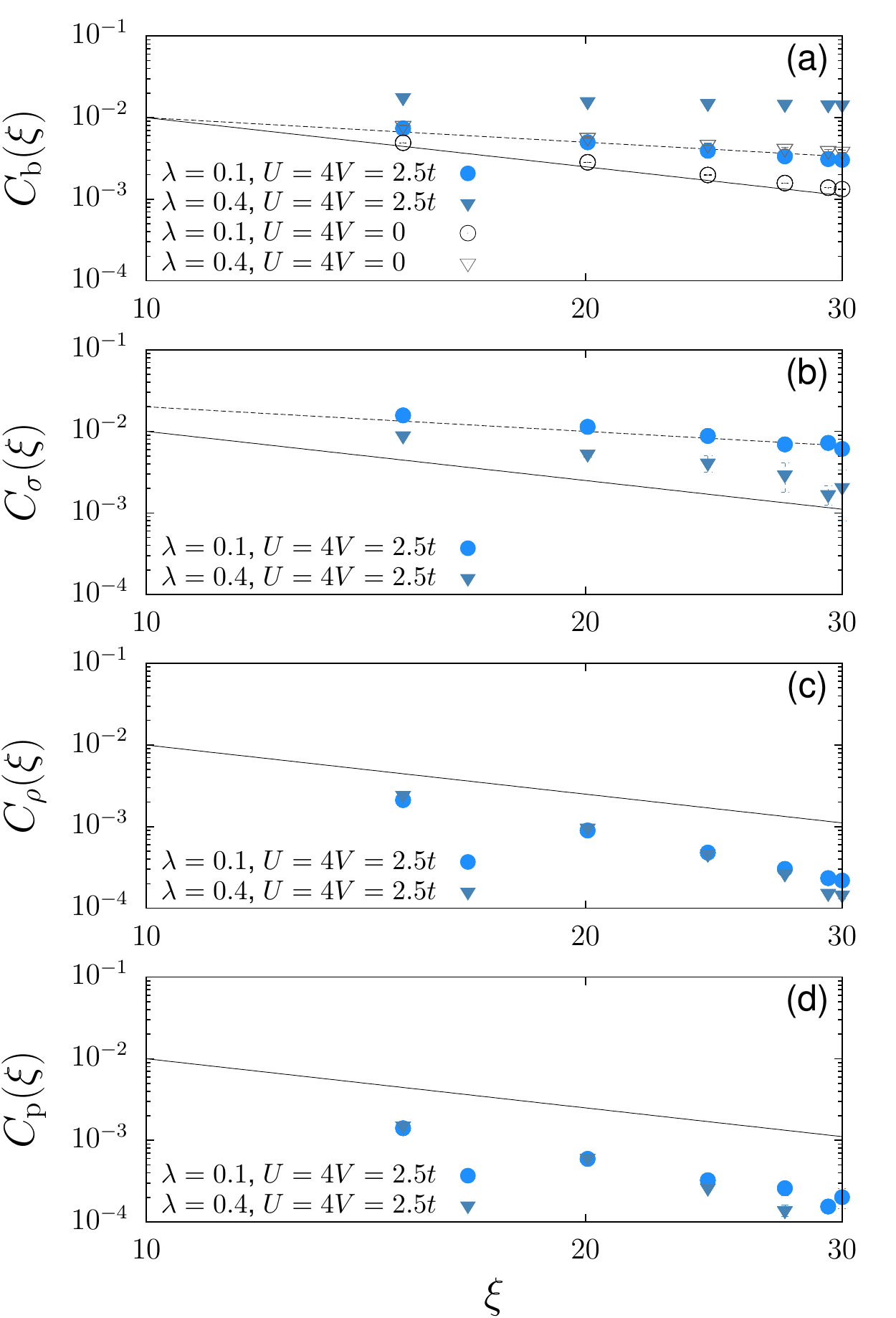}  
  \caption{\label{fig:correlations_UV_spinful} (Color online) (a) Bond,
    (b) spin, (c) charge, and (d) pairing correlations of the 
    optical SSH-$UV$ model as a function of $\xi= L\sin \left(\pi r/L\right)$. 
    The solid (dashed) line illustrates a $1/r^2$ (a $1/r$) decay.  Here, $\beta t
    =L=30$, $\om_0=0.5t$, $U=2.5t$, and $V=0.625t$.}
\end{figure}

Figure~\ref{fig:correlations_UV_spinful} shows the real-space correlation
functions for bond, spin, charge and pairing operators in the Mott
($\lambda=0.1$) and the Peierls phase ($\lambda=0.4$), respectively. Because
the system has a charge gap in both phases, the charge and pairing
correlations (see Ref.~\cite{Voit94} for the explicit Luttinger liquid
results) in general decay exponentially; the former because $K_\rho=0$ and
$A_{\rho}\to0$, similar to the spinless case, the latter because the
exponents diverge for $K_\rho=0$. Figures~\ref{fig:correlations_UV_spinful}(c) and
\ref{fig:correlations_UV_spinful}(d) reveal a strong suppression 
compatible with an exponential decay. For nonzero $U$ and $V$, the SSH-$UV$
model is no longer invariant under the modified particle-hole transformation,
and the degeneracy of spin and charge correlations is lifted.

The bond and spin correlations are more interesting. In a Luttinger liquid,
their $2\kF$ components decay as \cite{0022-3719-21-35-003}
\begin{align}\label{eq:correl:LL} \nonumber
  C^{2\kF}_\mathrm{b}(r) 
  &=
  \frac{A_\mathrm{b}}{r^{\Kr+\Ks^{\phantom{-1}}}}\cos(2\kF r)\,,  
  \\
  C_{\sigma}^{2\kF}(r)
  &=  
  \frac{A_\sigma}{r^{\Kr+\Ks^{-1}}}\cos(2\kF r) \,,
\end{align}
where we neglected potential logarithmic corrections due to marginal
operators. In the Mott phase with $K_\rho=0$ and $K_\sigma=1$ (corresponding
to a Luttinger liquid in the spin sector, or C0S1 in the notation of
Ref.~\cite{PhysRevB.53.12133}), Eq.~(\ref{eq:correl:LL}) predicts a $1/r$
decay, in good agreement with our results in
Fig.~\ref{fig:correlations_UV_spinful}(a) and
\ref{fig:correlations_UV_spinful}(b). While we cannot exclude logarithmic
corrections on the accessible system sizes, our results are consistent with a
larger amplitude for spin correlations than for bond correlations,
$A_\sigma>A_\text{b}$. In the Peierls phase ($K_\rho=K_\sigma=0$, C0S0), bond
correlations become long-ranged and spin correlations decay exponentially as
a result of the nonzero spin gap, see Fig.~\ref{fig:correlations_UV_spinful}.

Although finite values of $U$, $V$ give rise to a Mott phase for small enough
$\lambda$, it has been observed that $2\kF$ bond correlations are actually
enhanced by Coulomb interactions
\cite{PhysRevB.26.4278,PhysRevLett.51.292,PhysRevLett.51.296,PhysRevLett.62.1053}.
Our results in Fig.~\ref{fig:correlations_UV_spinful}(a) confirm such an
enhancement. Interestingly, previous work showed that the power-law exponent
of $2\kF$ bond correlations is independent of $U$ and $V$
\cite{PhysRevLett.62.1053}, namely equal to one in the Mott phase and zero in
the Peierls phase; the data in Fig.~\ref{fig:correlations_UV_spinful}(a) are
compatible with this result.  The enhancement of bond correlations is
therefore related to a larger amplitude $A_\text{b}$. While $A_\sigma$ grows
continuously with increasing $U$, $A_\text{b}$ has been found to have a
maximum near $U\sim4t$
\cite{PhysRevLett.51.292,PhysRevLett.51.296,PhysRevB.29.4230,PhysRevLett.62.1053}.

\begin{figure}[t]
  \includegraphics[width=0.45\textwidth]{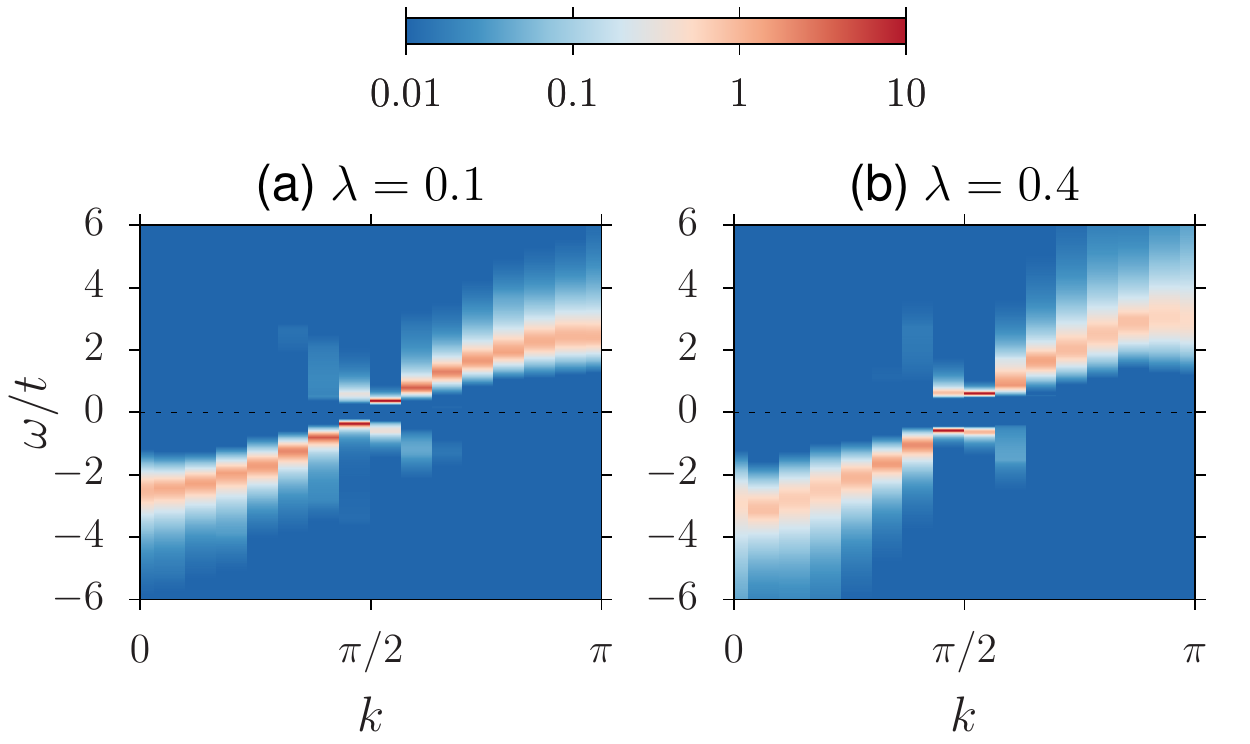}  
  \caption{\label{fig:dynamics_spinful_UV_akw} (Color online) Single-particle
    spectral function $A(k,\omega)$ of the optical SSH-$UV$ model. The dashed line indicates
    the Fermi level. Here, $\beta t =L=30$, $\om_0=0.5t$, $U=2.5t$,
    and $V=0.625t$.}
\end{figure}

Figure~\ref{fig:dynamics_spinful_UV_akw} shows the single-particle spectral
function for $\lambda=0.1$ [Mott phase,
Fig.~\ref{fig:dynamics_spinful_UV_akw}(a)] and $\lambda=0.4$ [Peierls phase,
Fig.~\ref{fig:dynamics_spinful_UV_akw}(b)]. In both cases, we find a gap at
the Fermi level that increases with increasing $\lambda$. Moreover, the SSH
electron-phonon coupling again gives rise to an enhanced bandwidth of the
cosine-like band.  For intermediate values near the Mott-Peierls transition
(not shown), we find no evidence for a closing of the gap, consistent with an
insulator-insulator transition without an intermediate metallic state.

\begin{figure}[t]
  \includegraphics[width=0.45\textwidth]{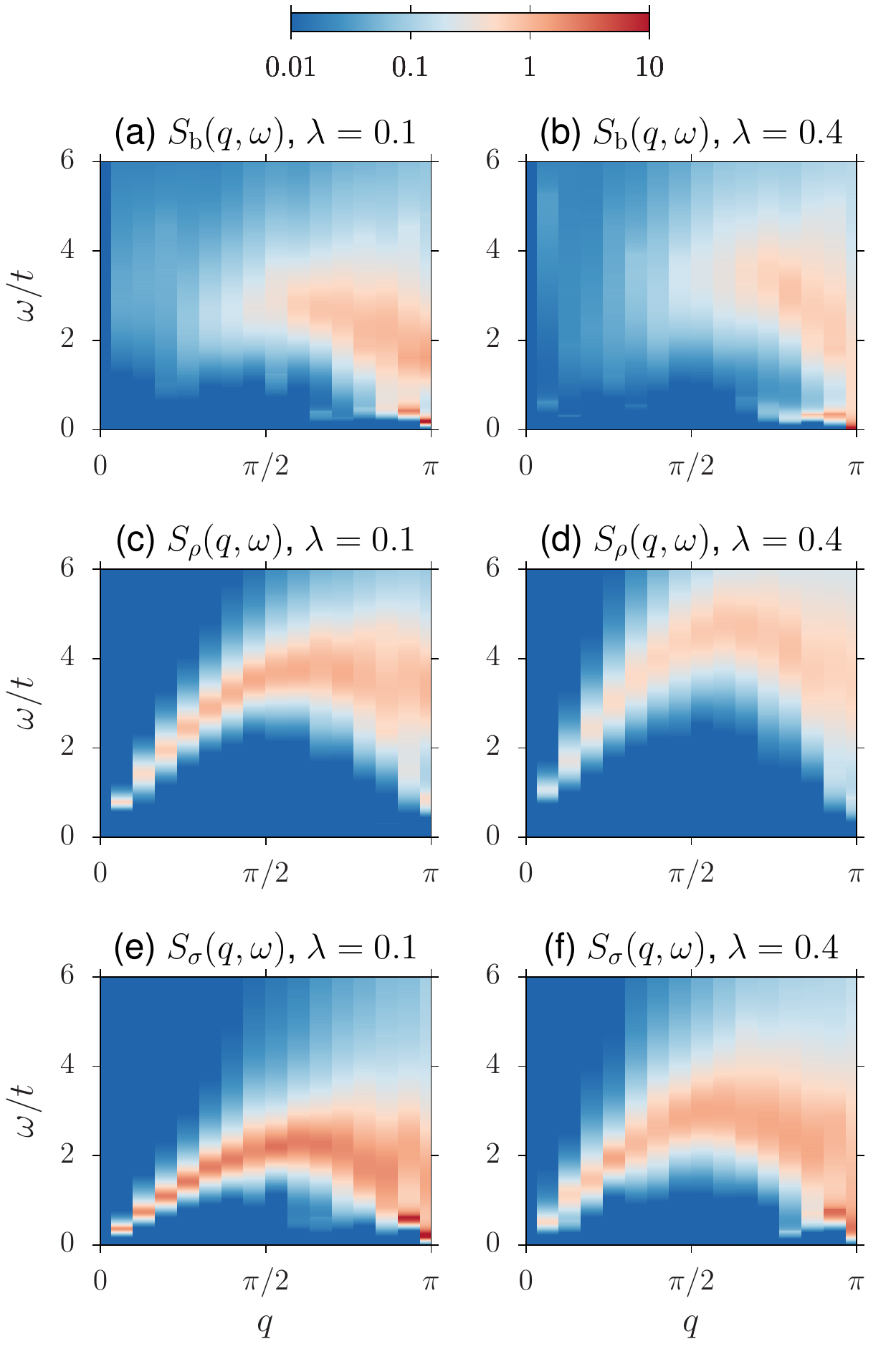}  
  \caption{\label{fig:dynamics_spinful_UV_lambda1} (Color online) (a)--(b) Dynamic
    bond, (c)--(d) dynamic charge, and (e)--(f) dynamic spin structure factor
    of the optical SSH-$UV$ model.  Here, $\beta t =L=30$, $\om_0=0.5t$, $U=2.5t$,
    and $V=0.625t$.}
\end{figure}

In Fig.~\ref{fig:dynamics_spinful_UV_lambda1}, we present the two-particle
excitation spectra in the Mott phase ($\lambda=0.1$) and the Peierls phase
($\lambda=0.4$). Whereas only a partial softening of the phonon mode at
$q=\pi$ is visible in the structure factor in the Mott phase
[Fig.~\ref{fig:dynamics_spinful_UV_lambda1}(a)], we observe a central peak at
$q=\pi$ in the Peierls phase
[Fig.~\ref{fig:dynamics_spinful_UV_lambda1}(b)]. The dynamic density
structure factor reveals a gap for long-wavelength excitations in both cases,
see Figs.~\ref{fig:dynamics_spinful_UV_lambda1}(c)
and~\ref{fig:dynamics_spinful_UV_lambda1}(d). Finally, the dynamic spin
structure factor in the Mott phase
[Fig.~\ref{fig:dynamics_spinful_UV_lambda1}(e)] has a gapless linear mode at
small $q$, and enhanced spectral weight at $q=\pi=2\kF$ reflecting the $1/r$
antiferromagnetic spin correlations. In the Peierls phase,
Fig.~\ref{fig:dynamics_spinful_UV_lambda1}(f), the weight at $q=\pi$ is
suppressed, but the spin gap at $q=0$ is much smaller than in
Fig.~\ref{fig:dynamics_spinful_lambda0.6}(b).

\section{Discussion}\label{sec:discussion}

\subsection{Agreement of SSH and optical SSH models}

Figure~\ref{fig:comparison} reveals a remarkable quantitative agreement
between results obtained for the SSH model~(\ref{eq:SSHq}) and the optical
SSH model~(\ref{eq:SSH0}). We have found this agreement to hold for all
parameters and observables considered. While theoretical arguments suggest
that only the phonon frequency at $q=2\kF$ is important for the low-energy
physics of the SSH model
\cite{PhysRevB.27.1680,PhysRevLett.60.2089,PhysRevB.29.4230}, deviations may
be expected at higher energies or smaller length scales, as typically probed
in numerical simulations. However, for the parameter sets considered, the
results are identical within error bars. This equivalence of the
models~(\ref{eq:SSHq}) and~(\ref{eq:SSH0}) is consistent with the excellent
agreement of the phase boundary for the optical SSH-$UV$ model from quantum
Monte Carlo simulations \cite{PhysRevB.67.245103} and that obtained with the
functional renormalization group \cite{Bakrim2015} for the SSH-$UV$ model
with acoustic phonons. On the other hand, a significant shift of the critical
value for the Mott-Peierls transition was reported in
Ref.~\cite{PhysRevB.83.195105}, although the values become almost identical
for small $U$ and $V$; however, these results were obtained for a model with
the original SSH coupling term.  Therefore, the optical limit does not
correspond to the optical SSH model considered here and in
Ref.~\cite{PhysRevB.67.245103}.

To understand the quantitative agreement between the SSH model and the
optical SSH model, it is useful to consider the relation of the two
models. The variables $\hat{Q}_i$ of the SSH model~(\ref{eq:SSHq}) describe
the longitudinal displacement of lattice site $i$ from its original position.
The hopping integral between sites $i$ and $i+1$ depends on the length
of the bond between these two sites, and hence on $\oQ_{i+1}-\oQ_{i}$. A
change of $\oQ_i$ also changes the bond between sites $i$ and $i-1$ because
the lattice displacements at neighboring sites are coupled; this coupling
gives rise to an acoustic phonon mode. In contrast, in the optical SSH
model~(\ref{eq:SSH0}), the variables $\hat{Q}_i$ describe the bond length
itself, that is, they correspond to the difference $\oQ_{i+1}-\oQ_{i}$ in
Eq.~(\ref{eq:SSHq}). Each of the $L$ bonds is modeled as a harmonic oscillator with displacement
$\hat{Q}_i$ and momentum $\hat{P}_i$. Neighboring bonds are considered to be
independent, as described by Einstein phonons. While the impact of this
approximation is {\it a priori} unclear, our numerical results suggest that
both models give identical results.

We attribute the quantitative agreement to the fact that at half-filling
($2\kF=\pi$), the $2\kF$ bond correlations---generic for Luttinger liquids
and Mott insulators [with a power-law decay as described by
Eqs.~(\ref{eq:correl:LLspinless}) and~(\ref{eq:correl:LL})] as well as for
Peierls insulators with long-range bond order---are equivalent to alternating
displacements of neighboring bonds. Because of the inherent preference for
$2\kF$ (alternating) bond correlations, it is not necessary to encode the
anticorrelation between the lengths of neighboring bonds into the
Hamiltonian. While the SSH and optical SSH models are not connected by a
canonical transformation, a change to bond variables $\hat{q}_i =
(\oQ_{i+1}-\oQ_i)/\sqrt{2}$ and conjugated bond momenta $\hat{p}_i$ in
Eq.~(\ref{eq:SSHq}) gives the optical SSH model~(\ref{eq:SSH0}) up to an
additional term of the form $\sum_i \hat{p}_i \hat{p}_{i-1}$ responsible for
the phonon dispersion. Neglecting this term amounts to replacing $\omega_q =
\om_\pi \sin (q/2)$ by $\om_q = \omega_\pi$. Because $\sin (q/2)$ has a maximum
at $q=\pi$, the first-order correction $\sim q$ vanishes for half-filling at
the dominant value $q=2\kF$ that determines the low-energy physics.  From
both arguments provided, we expect the models~(\ref{eq:SSHq})
and~(\ref{eq:SSH0}) to give different results away from half-filling where
$2\kF\neq \pi$.

\subsection{Lattice fluctuations and low-energy theory}

The question of the impact of quantum lattice fluctuations on the dimerized
Peierls state has been raised soon after the original work of SSH
\cite{PhysRevB.25.7789,PhysRevB.27.1680,PhysRevB.33.5141}. However, while
(thermal or quantum) fluctuations are in principle expected to destroy
long-range order for sufficiently weak electron-phonon interactions, the
Peierls state remains stable at $T=0$ in the spinful SSH model (although the
dimerization is reduced), as confirmed by renormalization group methods
\cite{PhysRevB.29.4230,Ba.Bo.07,PhysRevB.71.205113,Bakrim2015} and numerical
simulations
\cite{PhysRevB.27.1680,PhysRevB.67.245103,PhysRevB.73.045106,PhysRevB.83.195105}. In
this respect, the SSH model is hence quite different from the Holstein
model. In the latter, a metallic phase emerges from lattice fluctuations in
both the spinless and the spinful case
\cite{Hirsch83a,BuMKHa98,JeZhWh99}. The spinful SSH model hence constitutes a
case where Peierls' theorem is valid beyond the adiabatic limit.

For the SSH model, lattice fluctuations destroy long-range order if
additional electron-electron interactions are taken into account
\cite{PhysRevB.67.245103,PhysRevB.83.195105}. In this case, the critical
value $\lambda_c$ depends strongly on the phonon frequency
\cite{PhysRevB.67.245103}. However, a metallic state does not seem to exist.
Instead, the phase diagram in the $(\lambda,\omega_0)$ plane for fixed large
$U$ reflects the physics of the Heisenberg spin-phonon model, which is always
dimerized in the adiabatic limit, but has a gapless phase with critical spin
correlations for $\omega_0>0$ and $\lambda<\lambda_c$
\cite{PhysRevLett.83.195,PhysRevB.67.245103,Ci.Or.Gi.05,PhysRevB.74.214426}.
For large $U$, this phase corresponds to the Mott phase of the SSH model.

The absence of a metallic phase in the spinful SSH model was also predicted
from two-cutoff renormalization group calculations
\cite{PhysRevB.29.4230,Ba.Bo.07,PhysRevB.71.205113}. Because the forward
scattering matrix element $g_2$ is exactly zero, any repulsive umklapp
scattering $g_3$ is relevant. While this result agrees with numerical results
and the functional renormalization group, applications of the same method to
the Holstein model \cite{PhysRevB.29.4230,Ba.Bo.07,PhysRevB.71.205113} fail
to describe the extended metallic phase at weak coupling. The two-cutoff
renormalization group is hence not reliable in general. The likely reason why
it gives correct results for the SSH model is that the physics of the latter
at finite $\omega_0$ is essentially the same as for $\omega_0=\infty$
\cite{PhysRevB.27.1680}, so that retardation effects do not play a
significant role.

While a purely fermionic theory is in general not sufficient to capture the
physics of systems with a coupling to quantum phonons, a comparison with
bosonization and renormalization group results for fermionic models provides
valuable insights. Although the microscopic problem is rather different, the
metal-insulator transition of the spinless SSH model can be related to a
spinless Luttinger liquid that undergoes a Mott transition to a gapped $2\kF$
charge-density-wave phase. In fact, in the limit $\omega_0=\infty$, the
spinless SSH model maps to a model of spinless fermions with hopping $t$ and
nearest-neighbor repulsion $v=\lambda$, which exhibits a quantum phase
transition from a Luttinger liquid to a charge-density-wave insulator at
$v=2t$ (corresponding to $K_\rho=1/2$) \cite{Giamarchi}. Similarly, numerical
results for the spinless Holstein model---for which a strong-coupling picture
also leads to a model of spinless fermions with repulsion $v$---also support
a metal-insulator transition at $K_\rho=1/2$ \cite{Ej.Fe.09}.

The physics of the spinful SSH and SSH-$UV$ models may be related to
bosonization results for the extended Hubbard model with onsite interaction
$u$ and nearest-neighbor interaction $v$ \cite{PhysRevB.45.4027}. Here, $u$
and $v$ correspond to effective interactions that emerge from both
electron-electron and electron-phonon interactions. For the case $U=V=0$, the
existence of spin and charge gaps for any $\lambda>0$ may be related to the
fact that for $u=0$, umklapp and backscattering are relevant for any $v>0$,
giving rise to a fully gapped phase with long-range $2\kF$ charge
correlations (rather than a bond-order wave as in the SSH model; for the
extended Hubbard model, $2\kF$ bond correlations have the same exponent as
$2\kF$ charge correlations but a vanishing prefactor
\cite{PhysRevB.45.4027}). On the other hand, the Mott-Peierls transition
observed for nonzero Coulomb interactions $U$ and $V$ (with $U\gg V$) may be
understood in terms of the extended Hubbard model with $u>0$ and $v>0$. For
the latter, umklapp scattering is relevant for any $u>0$ and gaps out the
charge sector \cite{PhysRevB.45.4027}, while a transition to a spin-gapped
$2\kF$ charge-density-wave phase takes place near $u=2v$
\cite{PhysRevB.45.4027}. The Mott phase with gapless spin excitations that
exists for $u\gtrsim2v$ is identical to the Mott phase observed for
$\lambda<\lambda_c$ in the SSH-$UV$ model, whereas the fully gapped
charge-density-wave phase may be related to the dimerized Peierls phase.
(For the extended Hubbard model at weak to intermediate couplings
$u$ and $v$, two independent transitions for the spin and charge excitations
as well as an intermediate bond-ordered phase exist
\cite{doi:10.1143/JPSJ.68.3123,PhysRevB.65.155113,PhysRevLett.89.236401,PhysRevLett.91.089701,Sa.Ba.Ca.04,PhysRevLett.99.216403}.)

Although a metallic phase is absent in the SSH model, we note that an
extension of an argument by Voit for charge-density-wave insulators
\cite{Voit98} implies that a metallic phase with dominant bond correlations
can only be realized in Luther-Emery liquids with a gap for spin excitations,
because according to Eq.~(\ref{eq:correl:LL}) spin and bond correlations
decay with the same power-law exponent in a Luttinger liquid with
$K_\sigma=1$.

Although the bosonization results for the extended Hubbard model capture
several key aspects of the SSH-$UV$ model, the continuum field theory does
not account for the spontaneous breaking of a discrete Ising $Z_2$ symmetry
at the Peierls transition. Instead, it provides a theory of $U(1)$ charge and
spin phase fields that can be pinned when umklapp and/or backscattering terms
become relevant, leading to Kosterlitz-Thouless transitions and massive
excitations but no long-range order. Accordingly, the Mott phase has a charge
gap and critical spin correlations, but does not break any symmetries. In
contrast, the Peierls phase is characterized by charge and spin gaps and a
broken Ising symmetry reflected in the central peak visible in
Fig.~\ref{fig:dynamics_spinful_UV_lambda1}(b). The existence of a $U(1)\times
Z_2$ symmetry in the continuum limit has recently been illustrated for the
spinless Holstein model \cite{WeAsHo15I}.

\section{Summary}\label{sec:conclusions}

We used the CT-INT quantum Monte Carlo method to study SSH models with
quantum phonons. First, we demonstrated that the original SSH model with
acoustic phonons gives results that agree within statistical errors with
those for the optical SSH
model, a property that can be related to the inherent tendency toward $2\kF$
bond correlations. Next, we studied the Peierls metal-insulator transition of
the spinless SSH model in terms of real-space correlation functions and
excitation spectra. The results are consistent with a Kosterlitz-Thouless
quantum phase transition at a critical Luttinger parameter $K_\rho=1/2$. The
spectra reveal the expected Peierls gap, soliton signatures, and a central
peak related to long-range order. In the spinful case, we confirmed the
absence of a metallic phase by detecting a spin gap even at weak coupling,
and by demonstrating a symmetry-related degeneracy of spin and charge
correlations. The spectral functions reveal a gap for single-particle, charge
and spin excitations, and a central peak.  Finally, we studied the spinful
SSH model with additional electron-electron interactions appropriate for
polyacetylene.  Our results are consistent with a phase transition from a
Mott state with critical spin and bond correlations to a Peierls state with
long-range bond order and exponential decay of all other correlation
functions. Electron-electron repulsion was found to enhance bond
correlations. In contrast to the Holstein-Hubbard model, we found no
indications for metallic behavior.

\vspace*{0.5em}

{\begin{acknowledgments}%
We are grateful to the J\"ulich Supercomputing Centre for computer time, and
acknowledge financial support from the DFG Grants No. AS120/10-1 and
Ho 4489/3-1 (FOR 1807). We thank S. Ejima for providing us with benchmark
results, and C. Bourbonnais, H. Fehske, and E. Jeckelmann for helpful
discussions.
\end{acknowledgments}}

\end{document}